\documentclass[aps,prb,amsmath,amssymb,superscriptaddress,reprint,
]{revtex4-2}

\usepackage{graphicx}
\usepackage{dcolumn}
\usepackage{bm}
\usepackage{hyperref}
\hypersetup{colorlinks=true,allcolors=blue}

\usepackage{siunitx}
\usepackage{physics}

\usepackage{color}

\usepackage{mathptmx}

\begin{document}


\title{Valley-resolved Fano resonance in monolayer transition metal dichalcogenides\\ nanoribbons with attached stubs} 

\author{Hui-Ying Mu}	
\affiliation{Hebei Provincial Key Laboratory of Photoelectric Control on Surface and Interface, School of Science, Hebei University of Science and Technology, Shijiazhuang, Hebei 050018, China}

\author{Nie-Wei Wang}
\affiliation{Hebei Provincial Key Laboratory of Photoelectric Control on Surface and Interface, School of Science, Hebei University of Science and Technology, Shijiazhuang, Hebei 050018, China}

\author{Ying-Na Du}
\affiliation{Hebei Provincial Key Laboratory of Photoelectric Control on Surface and Interface, School of Science, Hebei University of Science and Technology, Shijiazhuang, Hebei 050018, China}

\author{Xing-Tao An}
\email[Correspondence to: ]{ anxt2005@163.com}
\affiliation{Hebei Provincial Key Laboratory of Photoelectric Control on Surface and Interface, School of Science, Hebei University of Science and Technology, Shijiazhuang, Hebei 050018, China}
\affiliation{Key Laboratory for Microstructural Material Physics of Hebei Province, School of Science, Yanshan University, Qinhuangdao 066004, P. R. China}

\author{Jian-Jun Liu}
\affiliation{Physics Department, Shijiazhuang University, Shijiazhuang 050035, China}

\date{\today}

\begin{abstract}
Valley degree of freedom besides spin is a promising candidate as a carrier of information. Spintronics has come a long way and spin modulation can be realized by quantum interference and spin-orbit coupling effect. However, the control of valley degree of freedom using quantum interference is still a problem to be explored. Here we discover a mechanism of producing valley polarization in a monolayer transition metal dichalcogenides nanoribbon with attached stubs, in which valley-resolved Fano resonance are formed due to the quantum interference of intervalley backscattering. When the quantum interference occurs between the localized states at the edge of the stubs and the continuous channels in the nanoribbon, the transmission dips of Fano effect is valley-polarized. As the number of stubs increases, the valley-polarized transmission dips will split and valley-resolved minigaps are formed by Fano resonance with intervalley backscattering in stub superlattice. When the electron incident energy is in these valley-resolved gaps of the superlattice, even with several stubs, the transmission can have a significant valley polarization. Our finding points to an opportunity to realize valley functionalities by quantum interference
\end{abstract}

\maketitle

\sloppy

\section{\label{sec:Inro}Introduction}

Fano resonance arises from the quantum interference between two interfering configurations, one directly through the continuum states and the other through a discrete level \cite{x1}. The constructive and destructive interferences of the two configurations give rise to a characteristic asymmetric line shape in the spectrum. Although the Fano resonance is established in spectroscopy, it has been observed as a ubiquitous phenomenon in a wide variety of physical process with coexisting discrete and continuum states, particularly electronic transport in low-dimensional nanostructure \cite{x2,x3,x4,x5,x6,x7,x8,x9,x10,x11,x12,x13,x14,x15}. Both experimental and theoretical investigations have indicated that Fano resonance has potential application in spintronics \cite{x7,x8,x9,x10} and optoelectronics devices \cite{x16,x17}.

Fano resonance also occurs in monolayer materials, such as graphene and monolayer transition metal dichalcogenides (MTMDs) \cite{x6,x11,x18,x19}. The energy dispersion of electrons in these materials usually has a pair of degenerate minima located at well-separated momentum space points, known as valleys. The valley degree of freedom besides spin provides a feasible way to design information-storage or information-processing devices. Similar to the control of the spin polarization in spintronics, the manipulation of valley polarization configurations, i.e., unequal population distribution among the degenerate valleys, is one of the key elements for valleytronics devices. The valley polarization has been widely studied by using edges \cite{x20}, doping \cite{x21}, defects \cite{x22,x23} lattice strains \cite{x24,x25,x26,x27,x28}, intervalley scattering \cite{x29,x30}, or valley-dependent trigonal warping of the dispersions \cite{x31}. Meanwhile, the spin-like properties of the valley, including the valley Hall effect \cite{x32,x33,x34}, the valley magnetic response \cite{x21,x35,x36,x37,x38,x39,x40,x41,x42}, and the valley optical selection rules \cite{x43,x44}, allow its manipulation similar to the spin controls. A large spin polarization can be generated in semiconductor nanostructures those involves both the spin properties and the quantum interference effects related to spin-dependent Fano resonance. However, it is still a pertinent challenge to realize valley polarization in monolayer honeycomb lattice materials using valley quantum interference effects, such as valley-dependent Fano resonance.

Here we discover that, at a MTMDs nanoribbon with an attached stub, the Fano resonance, i.e., the dip in the transmission spectrum caused by the quantum interference between the localized states of the stub and the continuum states in the nanoribbon, is split into two valley-resolved dips. This valley selectivity is made possible by the intervalley backscattering induced by the localized states in the stub. When the number of stubs attached to the nanoribbon increases, each valley-resolved Fano dip will further split into multiple dips. Therefore, the valley-resolved minigaps will be formed for stub superlattice structures. A sizeable valley polarization can be obtained when the electron incident energy is in these valley-resolved gaps even for the nanoribbon attached several stubs.

\begin{figure}
\includegraphics[width=8.5cm]{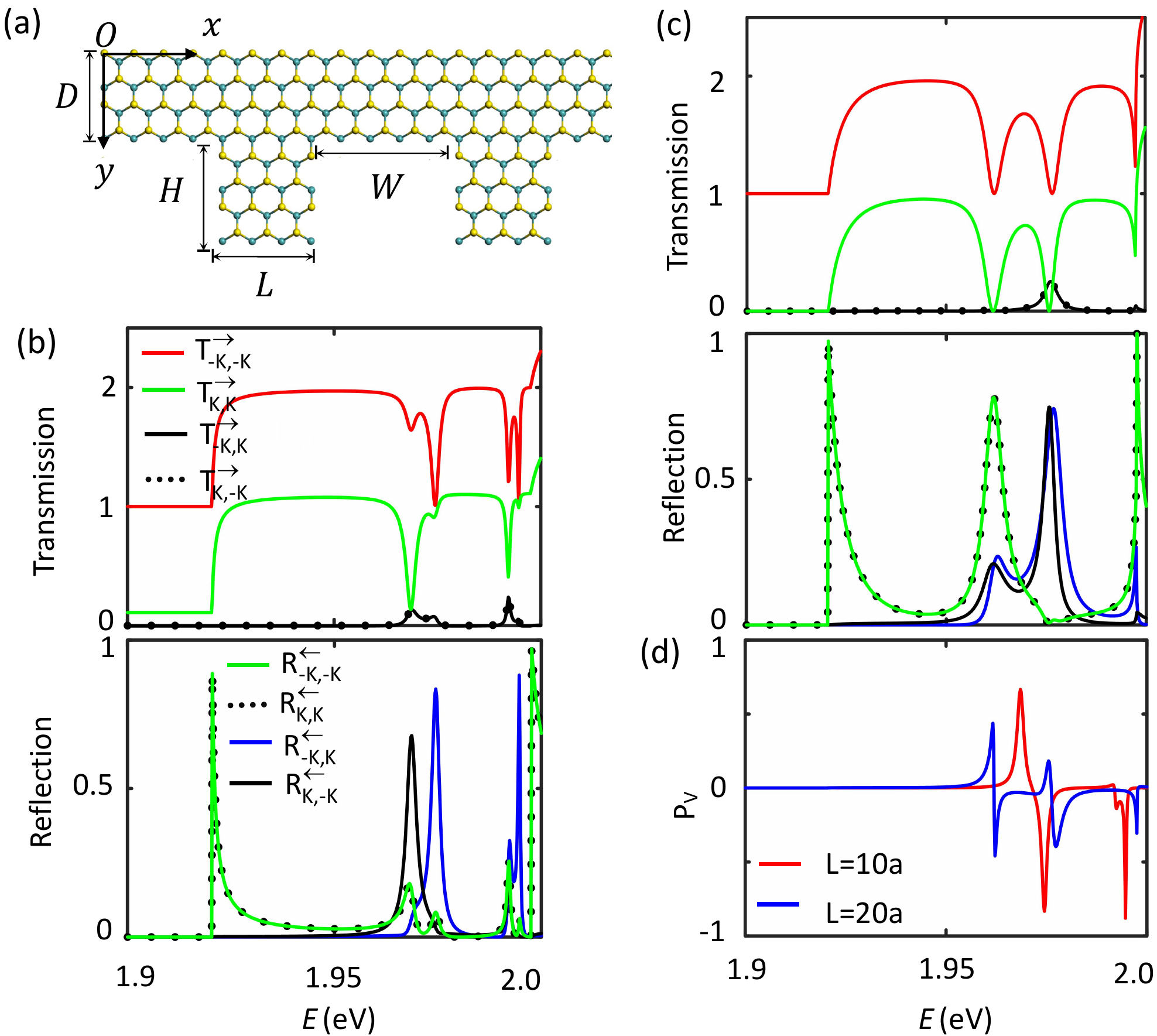}
\caption{\label{fig:a1}Valley-resolved Fano resonance as electron transmission through the MTMDs nanoribbon with an attached stub. (a) Schematic of nanoribbons with attached stubs connected to MTMDs leads on the two sides. (b) Valley-resolved transmission and reflection coefficients as functions of the incident energy, with $L=10a$, $a$ being the lattice constant of MTMDs. (c) Valley-resolved transmission and reflection coefficients for $L=20a$. (d) Valley polarization $(P_{v})$ of the transmission as a function of the incident energy for the two configurations. All calculations here use parameters $D=10\sqrt3a$, $H=4\sqrt3a$ (c.f. Fig.~\ref{fig:a1}(a)).
}
\end{figure}

\section{\label{sec:theor}MODEL AND METHODS}

Let us consider a MTMDs zigzag nanoribbon with attached stubs of length $L$ and width $H$ (Fig.~\ref{fig:a1}(a)). The width of the nanoribbon is $D$ and the spacing of the stubs is $W$. In MTMDs, the conduction edges at the $\pm$$K$ valleys are contributed predominantly by the three metal d-orbitals \cite{x45}: $d_{z^{2}}$, $d_{xy}$, $d_{x^{2}-y^{2}}$. For a quantitative characterization of the valley-polarized transport and superlattice minibands, with the Fano interference of intervalley backscattering, we have calculated the valley-dependent transport properties and energy dispersion using the tight-binding model constructed with the three orbitals that describes well the band edge electrons \cite{x45},

\begin{equation}\label{eqi1}
H=\sum_{i} \sum_{\mu} \varepsilon_{i \mu} c_{i \mu}^{\dagger} c_{i \mu}+\sum_{<i, j>} \sum_{\mu v} t_{i \mu, j v} c_{i \mu}^{\dagger} c_{j v} .
\end{equation}

\noindent Here $\epsilon_{i\mu}$ and $c^{\dagger}_{i\mu}$ are, respectively, the on-site energy and the creation operator for the electron on orbital $\mu$ at metal site $i$, the sums $<i,j>$ run over all pairs of nearest-neighbor metal sites, and $t_{i\mu,i\nu}$ are the hopping terms fitted from first principles band structures \cite{x45}.

Consider the nanoribbon with stubs connected to the outer world by left and right semi-infinite pristine MTMDs leads, the valley-resolved transport properties of the system with the tight-binding Hamiltonian in Eq. (\ref{eqi1}), are calculated using a recursive Green's function technique \cite{x46} in Appendix~\ref{RG}.

\section{RESULTS AND DISCUSSION}\label{sec3}

As examples, we numerically demonstrate the sizable valley polarization effects driven by Fano resonance of intervalley backscattering in monolayer MoS$_{2}$ nanoribbons with stubs. The widthes of the nanoribbon and the stub are set as $D=10\sqrt3a$ and $H=4\sqrt3a$, respectively, $a$ being the lattice constant of MoS$_{2}$. Fig.~\ref{fig:a1}(b) shows the calculated valley-conserved and valley-flip transmission and reflection coefficients for the nanoribbon attached one stub with $L=10a$. The two valley-conserved reflection coefficients $R_{K,K}^\leftarrow$ and $R_{-K,-K}^\leftarrow$ are always identical, as they correspond to intravalley backscatterings with the same momentum transfers. The two valley-flip transmission coefficients are equal, i.e. $T_{K,-K}^\rightarrow=T_{-K,K}^\rightarrow$, as the two scattering channels are conjugate of each other \cite{x29}.  Both of the valley-conserved transmission coefficients, $T_{K,K}^\rightarrow$ and $T_{-K,-K}^\rightarrow$ in Fig.~\ref{fig:a1}(b), exhibit a wide plateau and a series of dips on the plateau. These dips arise from the Fano-type interference between the continuous states propagating in the main channel and a bound state formed in the stub. This type of resonance is called as the structure induced Fano resonance. The peaks of valley-flip reflection coefficients, in Fig.~\ref{fig:a1}(b), indicate that the Fano resonance here is a quantum destructive interferences behavior with strong intervalley backscattering. The valley-resolved Fano resonance of transmission coefficients in Fig.~\ref{fig:a1}(b) is caused by the combined effect of intervalley backscattering and quantum interference.

The charge current passing through the nanoribbon with a stub is accompanied by a valley-polarized flow when the electron incident energy is set at the Fano resonant position. In order to study the valley polarization caused by Fano resonance with intervalley backscattering, we do not consider the effect of the edge state, because the valley-dependent transport properties of the edge state are obvious. At the upper boundary of the nanoribbon, the electrons in $-K$ valley propagate steadily to the right lead and will not be affected by the stub \cite{Sekine102}. At the bottom of the conduction band, the edge states propagating to the right belong only to the $-K$  valley. In any low energy region, the contribution of edge states to transmission is always 1. Here, what we are concerned with is the contribution of bulk states to valley polarization, so we remove the edge state contribution. The valley polarization can be defined as  $P_{\nu}\equiv\frac{1}{T}(T_{K,K}^\rightarrow+T_{-K,K}^\rightarrow-{\widetilde{T}_{-K,-K}^\rightarrow}-T_{K,-K}^\rightarrow)$, where $\widetilde{T}_{-K,-K}^\rightarrow=T_{-K,-K}^\rightarrow-1$ and $T=T_{K,K}^\rightarrow+T_{-K,K}^\rightarrow+\widetilde{T}_{-K,-K}^\rightarrow+T_{K,-K}^\rightarrow$ remove the edge state contribution. Fig.~\ref{fig:a1}(d) plots the valley polarization as a function of the incident energy for $L=10 a$ and $L=20 a$. A pronounced valley polarization almost reaching $\sim85\%$ can be obtained in the nanoribbons with a short stub, for example $L=10a$, in which the intervalley backscattering results significant valley-resolved Fano resonance. Moreover, the direction of valley polarization can be controlled by the incident energy.

We also examined the effect of the stub size on the Fano resonance. When the size of the stub increases, two kinds of Fano resonances appear on the transmission spectra plateau, as shown in Fig.~\ref{fig:a1}(c). It can be seen from the reflection coefficients that one is mainly caused by intravalley backscattering and the other is induced by intervalley backscattering. The valley splitting effect of Fano resonance will be suppressed when the size of the stub increases. The valley-resolved Fano resonance positions are difficult to distinguish in terms of the incident energy. In this case, valley polarization generates near the Fano resonance energy positions, but the strength of the valley polarization is reduced, as shown in Fig.~\ref{fig:a1}(d).

\begin{figure}
\centering
\includegraphics[width=8.5cm]{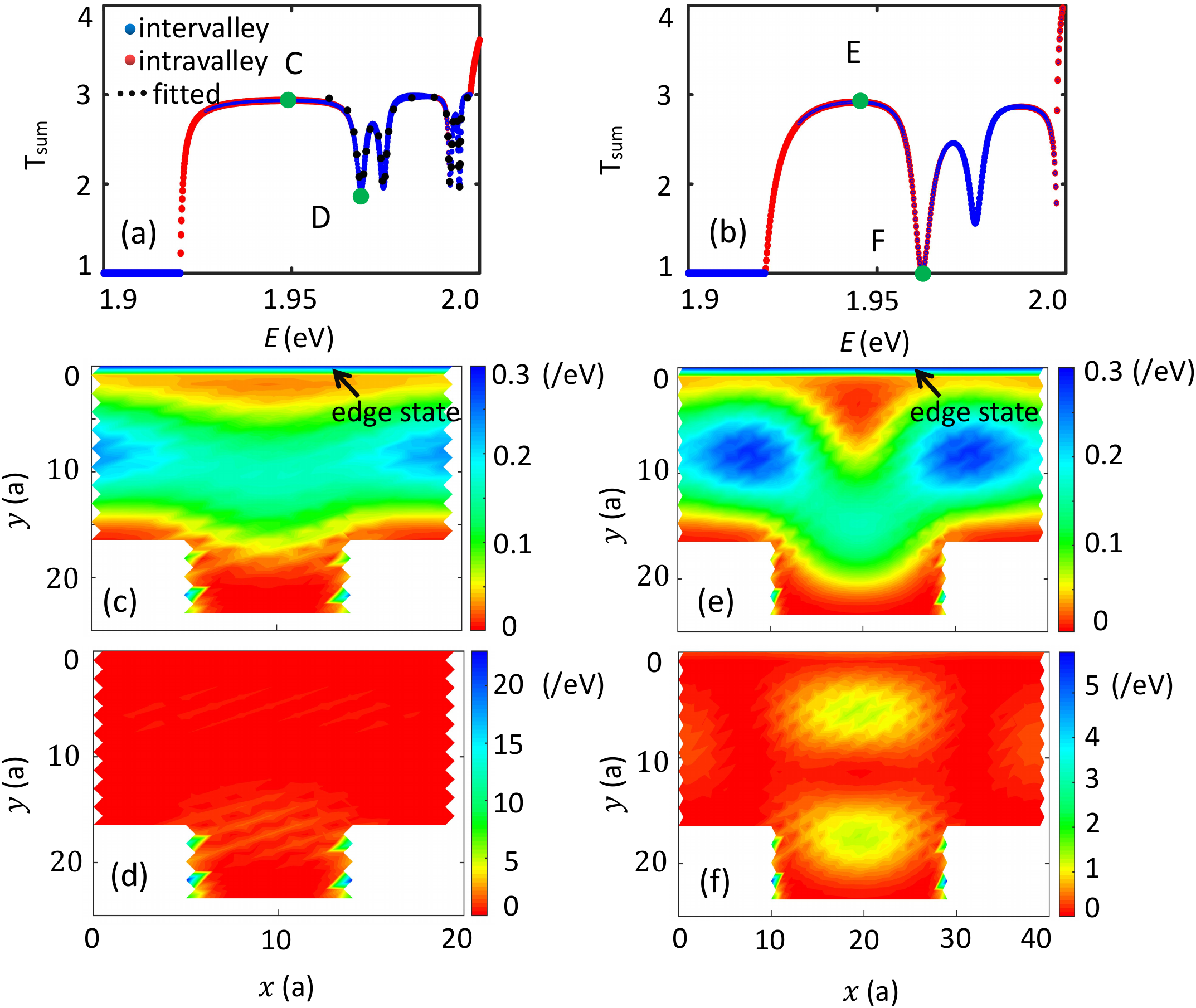}
\caption{\label{fig:a2}Local density of state of the MTMDs nanoribbons with a stub. (a) The total transmission and fits (black dotted line) as a function of the incident energy for $L=10a$. Blue and red dots represent intervalley and intravalley backscattering respectively, and their  size of the symbol represents their contribution to the total transmission. (b) The total transmission for $L=20a$. (c-d) Profiles of the local density of state as the incident energies are taken at the positions marked by the corresponding four green points (C, D, E and F) as shown in (a) and (b). Black arrows indicate edge states in (c) and (e).
}
\end{figure}

In order to offer the most general guides for achieving the best result of the valley polarization, we can discuss the location of the dips in the transmission spectra. It depends on the energy levels in the stubs. An analytical expression for the spectral intensity was first proposed and later applied to QD-AB-ring \cite{x3} and T-shaped quantum waveguides \cite{Klaiman},
\begin{equation}\label{eqi13}
T(\varepsilon) \propto \frac{(\varepsilon+q)^{2}}{\varepsilon^{2}+1},
\end{equation}

\noindent with $\varepsilon=\frac{E-E_{0}}{\Gamma / 2}$, where $E_{0}$ and $\Gamma$ are the energy position and width of the resonance state, respectively. The Fano parameter $q$ is a measure of the coupling strength between the continuum state and the resonance state. The Fano parameter $q$ selects from a symmetric peak $(q=\infty)$ or dip $(q=0)$, or a dip to the left $(q\textgreater0)$ or right $(q\textless0)$ of a peak. Here, we can set $q=0$ and roughly fit the dips of the transmission for the valley-resolved Fano resonance in Fig.~\ref{fig:a2}(a). There are two kinds of intervalley scatterings with distinct momentum transfers \cite{x29}. In the first Born-approximation, $R_{K,-K}^\leftarrow$ and $R_{-K,K}^\leftarrow$ simply correspond to different Fourier components of the scattering potential at $2K+2q_{F}$   and $2K-2q_{F}$ respectively, where $q_{F}$ is the Fermi wavevector. Due to the quantum size effect, two kinds of intervalley scatterings lead to valley-resolved energy levels in the stub. According to Eq. (\ref{eqi13}), the valley-resolved energy levels distinguish the transmission in the two valleys at the energies of Fano resonance.

Figs.~\ref{fig:a2}(a)-(b) plot the total transmission, $T_{sum}=T_{K,K}^\rightarrow+T_{-K,K}^\rightarrow+T_{-K,-K}^\rightarrow+T_{K,-K}^\rightarrow$, as a function of the incident energy for $L=10a$ and $L=20a$, respectively. In the bulk energy gap below the bottom of the conduction band, the total transmission $T_{sum}=1$ is the contribution of the edge state. For Fano resonance with intervalley backscattering, such as point D in Fig.~\ref{fig:a2}(a), the quantum interference is not completely destructive and one of the valleys is retained in the total transmission. For F point in Fig.~\ref{fig:a2}(b), Fano resonance with intravalley backscattering is completely destructive quantum interference. It is pointed out that the Fano resonances are contributed by both intervalley and intravalley backscattering. For some resonance dips, one of them plays the dominate role. For example, the F point in Fig.~\ref{fig:a2}(b) corresponds to a resonance mainly due to intravalley backscattering, while the D point in Fig.~\ref{fig:a2}(a) is a resonance mostly contributed by intervalley backscattering. We have superimposed the relative contributions of intravalley backscattering and intervalley backscattering on the transmission spectrum to identify the two backscattering processes more easily, as shown in figure Figs.~\ref{fig:a2}(a)-(b).

In order to understand the localized state distribution in the stub which causes the valley splitting effect of Fano resonance in detail, in Figs.~\ref{fig:a2}(c)-(f), we plot the profiles of the local state density of the system at different electron incident energies denoted by the four green points in Figs.~\ref{fig:a2}(a)-(b). The edge state at the upper boundary of the nanoribbon will not be destroyed by the stub at any incident energy shown in Figs.~\ref{fig:a2}(c)-(f). When the electron incident energy is set in the transmission plateau region, there is no localized states in the stub and the electron is perfectly transmitted through the continuous states in the nanoribbon, as Fig.~\ref{fig:a2}(c) illustrates. As can be seen from the profile of the local state density at point E in Fig.~\ref{fig:a2}(e), when the stub length increases, the continuous channels in the nanoribbon shift to the stub, but the electron is still not localized by the stub.

\begin{figure}
\centering
\includegraphics[width=8.5cm]{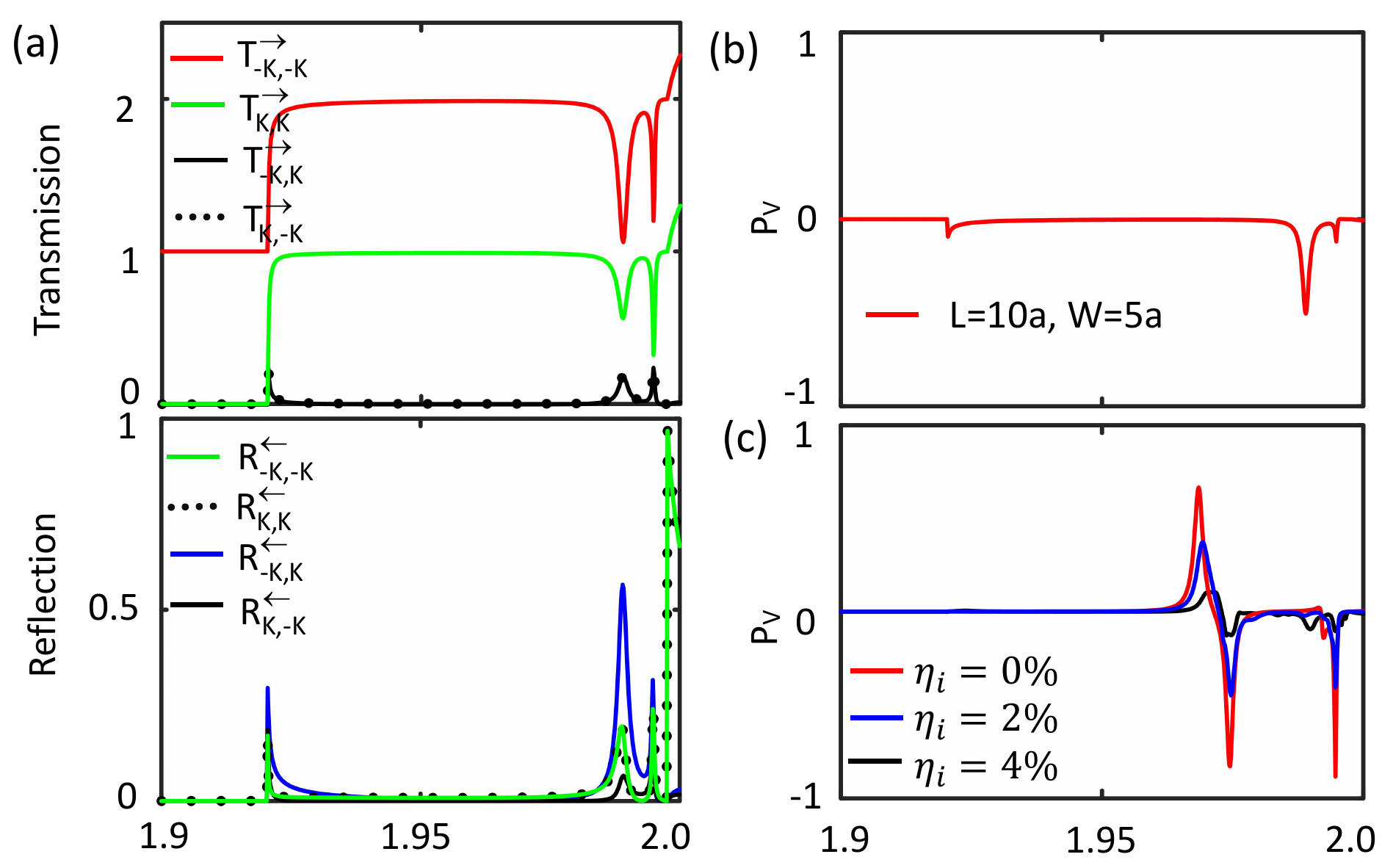}
\caption{\label{fig:aa2} The effects of different edges and defects on the valley-resolved Fano resonance. (a) Valley-resolved transmission and reflection coefficients as functions of the incident energy for a stub with left and right zigzag edges, with $L=10a$ and $W=5a$. (b) Valley polarization as a function of the incident energy for a stub with zigzag edges. (c) Valley polarization as a function of the incident energy for various defects density in the stub, and $\eta_{i}$ is the defects density.
}
\end{figure}

At the intervalley-backscattering Fano resonant energy positions, such as point D, the electrons are sharply localized on the left and right boundaries of the stub. The distribution range of the local states can be compared with the lattice constant of the MTMDs, so there is a significant intervalley backscattering effect during electron transmission. It is the localized states on the armchair boundary of the stub that gives rise to valley-polarized Fano resonance. As shown in Fig.~\ref{fig:a2}(f), for the incident energy corresponding to the Fano resonant mainly induced by intravalley backscattering, the local state density is highly localized inside the stub region and forms two isolated islands. In this situation, standing waves are formed by the interference between the electron waves reflected from the walls of the stub and those in the nanoribbon.

Fig.~\ref{fig:aa2}(a) and Fig.~\ref{fig:aa2}(b) show the Fano resonance and the resultant valley polarization when the left and right boundaries of the stub are zigzag edges. It is found that the valley splitting effect of Fano resonance is suppressed in this case. The reason is that the intervalley scattering is not significant in the stub with zigzag edges so as to the valley-resolved energy levels splitting is weakened. Therefore, the armchair edge of the stub is one of the important factors to obtain the high valley polarization. Fig.~\ref{fig:aa2}(c) shows the valley polarization as a function of the incident energy for various defects density $\eta_{i}$ in the stub. Although the location of the dips remained the same, the strength of the dips became weaker. Valley polarization decreases with the increase of defects density in stub area.

\begin{figure}
\centering
\includegraphics[width=8.5cm]{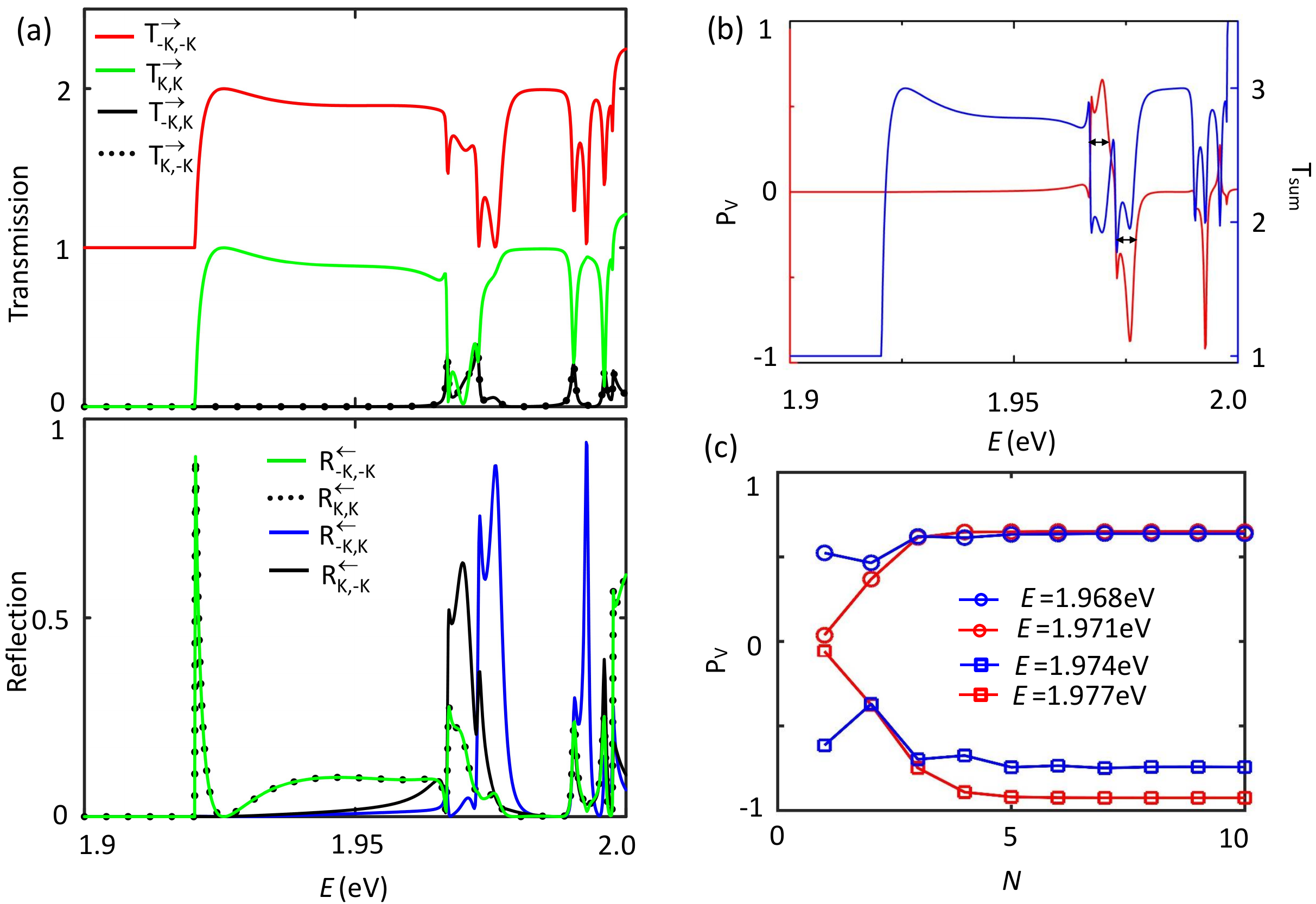}
\caption{\label{fig:a3}Valley-polarized transmission through the MTMDs nanoribbon with multi-stubs. (a) Valley-resolved transmission and reflection coefficients as functions of the incident energy for two stubs, with $L=10a$ and $W=5a$, (b) The corresponding valley polarization and the total transmission as functions of the incidence energy. (c) Valley polarization as a function of the number of the stub for various incident energies in the region indicated by double-headed arrows in (b).
}
\end{figure}

The sharpness of the valley-polarized Fano resonances is limited to valley filtering applications, so we also studied the effect of the number of stubs on the valley-polarized transport properties. Fig.~\ref{fig:a3}(a) shows the calculated valley-conserved and valley-flip transmission and reflection coefficients for a two-stub nanoribbon with $L=10a$, and $W=5a$. Each valley-polarized Fano resonance dip is split into two, which broadens the range of the incident energy for valley polarization. Each valley polarization peak of the nanoribbon with one stub becomes two peaks now and the valley polarization maintains a high intensity in a certain incident energy range, as shown in Fig.~\ref{fig:a3}(b). If the number of the stubs attached to the nanoribbon continues to increase, the valley-polarized Fano resonance dips and the valley polarization peaks split further into more dips and peaks. Fig.~\ref{fig:a3}(c) shows the valley polarization as a function of the number of the stubs when the incident energy is in the energy range indicated by double-headed arrows in Fig.~\ref{fig:a3}(b). Large valley polarization is obtained over a wide energy range, and remarkably, by just using four stubs, the valley polarization can already reach $60\%\sim90\%$ (c.f. Fig.~\ref{fig:a3}(c))

\begin{figure*}
\centering
\includegraphics[width=16cm]{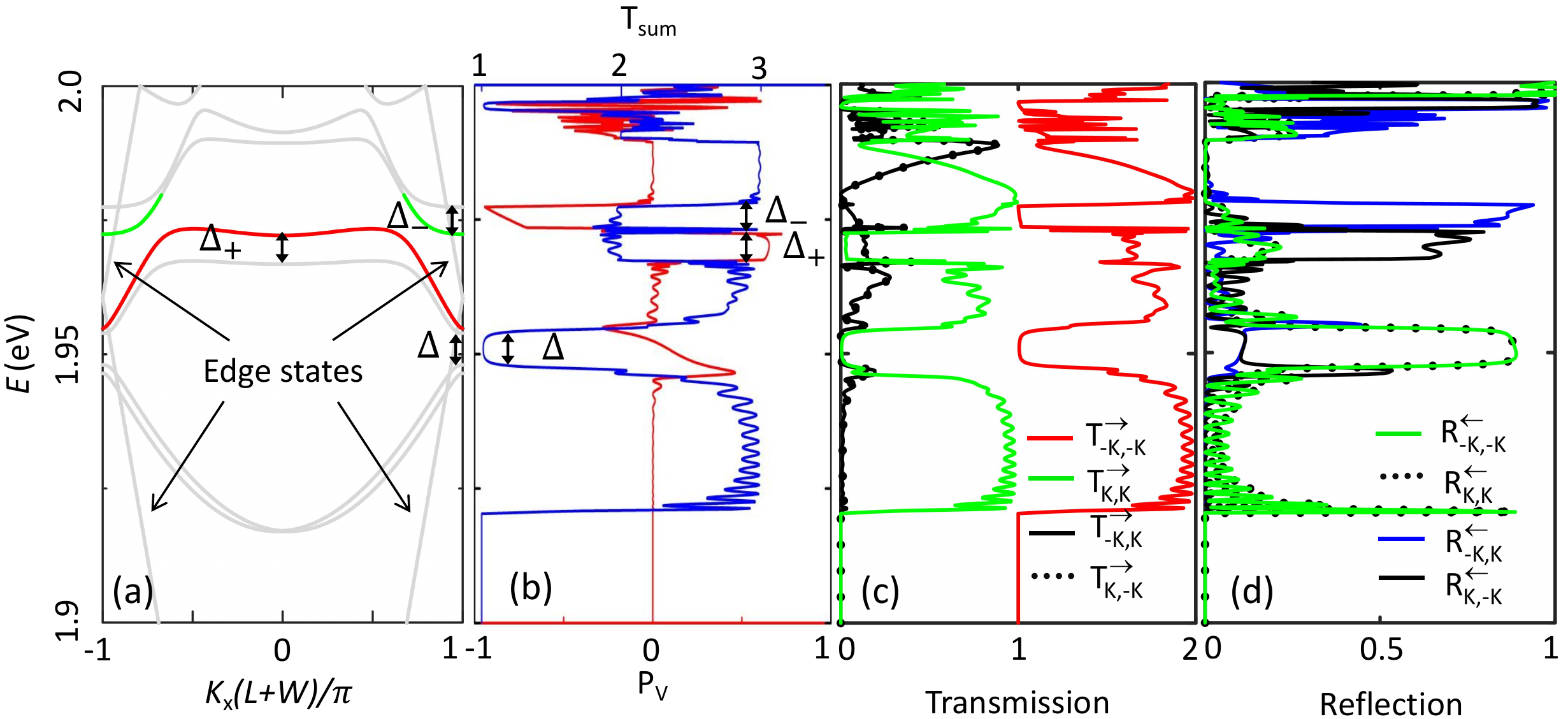}
\caption{\label{fig:a4}Valley filtering performance for the stub superlattice. (a) An example of superlattice miniband dispersion, with  $L=10a$ and $W=5a$. The lines indicated by the arrows are the edge states of the system. In the case of zone folding, intravalley backscattering causes a bulk minigaps $\Delta$ at the boundary of the mini-zone. The intervalley backscattering gives rise to two valley-resolved gaps $\Delta_{+}$ and $\Delta_{-}$ at the center and boundary of the mini-zone, respectively. The upper one of the two bands split by $\Delta_{+}$ is responsible to the transmission of the $-K$ valley state, which are colored in red. The lower one of the two bands split by $\Delta_{-}$ is responsible to the transmission of the $K$ valley state, which are colored in green.  (b) Valley polarization and transmission as functions of the incidence energy. (c) Valley-resolved transmission coefficients. (d) Valley-resolved reflection coefficients. The number of superlattice periods $N=20$ is used in (b-d).
}
\end{figure*}

According to the superlattice transport theory, as long as the distance between the stubs is not too far and the wave functions in the adjacent stubs can overlap, the nanoribbon with multiple stubs will have the properties of miniband transport, that is, the system will have valley-resolved minibands and minigaps. Fig.~\ref{fig:a4}(a) shows an example of the stub superlattice energy bands, with $L=10a$ and $W=5a$, where the length of the supercell is $L+W$. In the superlattice, the edge states still exist perfectly (indicated by the arrows in Fig.~\ref{fig:a4}(a)), and zone folding does not open the energy gap in the edge states, because the edge states can not be reflected by the stubs. Our calculation finds that the strong intravalley backscattering shown in Fig.~\ref{fig:a4}(d), leads to a sizable bulk minigap $\Delta$ at the boundary of the mini-zone.

Besides the bulk minigap $\Delta$, there are two valley-resolved minigaps $\Delta_{+}$ and $\Delta_{-}$ at the center and boundary of the mini-zone, respectively.  In the neighborhood of $\Delta_{+}$ and $\Delta_{-}$, multiple intervalley backscattering by the stubs gives rise to closely spaced Fano resonance dips which will develop into minigaps in infinite period superlattice. Within the gaps $\Delta_{+}$ and $\Delta_{-}$, the minibands are valley polarized. The Fano resonance with intervalley backscattering makes possible energy windows for valley-polarized transport in the superlattice. Figs.~\ref{fig:a4}(b)-(c) shows the valley polarization and transmission as functions of the incident energy, for a 20-stub superlattice with $L=10a$ and $W=5a$. As the periods increase, the Fano resonance dips in transmission are further split groups of valley-resolved dips. These dips evolve into continuous valley-resolved minigaps in the limit of a superlattice. This leads to a pronounced valley polarization exceeding $90\%$, with a total transmission $T_{\text {sum }} \sim 2$ (Fig.~\ref{fig:a4}(b)). If the valley polarization defined by the conductance, the transmission coefficient should be weighted by the Fermi distributions of the leads and integrated over energy. The valley imbalance regarding the carrier densities achieved by the valley filter may be significantly diluted due to the valley-resolved gaps are narrow and adjacent to each other.
\section{Conclusions} \label{sec4}

In summary, we have shown that a remarkable valley filtering can be realized in a MTMDs nanoribbon with attached stubs due to the Fano resonance with the intervalley backscattering. The edge states on the boundaries of the stub and the bulk localized states in the stubs cause respectively the intervalley and intravalley-backscattering Fano resonance. It is the intervalley-backscattering Fano resonance that results a significant valley polarization. The valley-resolved Fano resonance will split when the number of stubs increases and the valley-resolved minigaps are formed at the supercell Brillouin zone boundary and center for the stub superlattice structure. The transmission has nearly perfect valley polarization in these gaps with alternating valley polarity. This broadens the valley-polarized energy region and makes possible control of the valley filtering functionality by electrostatic control. Moreover, a valley polarization in a wide energy region can be generated in nanoribbons with just several stubs. These results point to an unexpected but exciting opportunity to build valley functionality by quantum interference in MTMDs nanostructure.

\begin{acknowledgments}
This work was supported by the National Natural Science Foundation of China (No.12074096) and Hebei Province Natural Science Foundation of China (No. A2021208013).
\end{acknowledgments}

\appendix

\section{RECURSIVE GREEN'S FUNCTION TECHNIQUE}
\label{RG}

In the main text, we used the recursive Green's function method, which is very convenient in solving quantum transport. This method is utilized in the calculation of valley pump of electrons or holes at nonmagnetic disorders \cite{x29}. Here we derive the recursive Green's function method. We assume  each  unit cell with an index $i$, the equation of motion can be written as

\begin{equation}\label{eqi2}
-\left(E \boldsymbol{I}-\boldsymbol{H}_{i}\right) \boldsymbol{C}_{i}+\boldsymbol{H}_{i, i-1} \boldsymbol{C}_{i-1}+\boldsymbol{H}_{i, i+1} \boldsymbol{C}_{i+1}=0 ,
\end{equation}

\noindent where $\boldsymbol{C}_{i}$ is a vector describing the wave-function coefficients on all sites and orbits of unit cell $i$ within three-band tight-binding model. The matrices $\boldsymbol{H}_{i}$ and $\boldsymbol{H}_{i,i+1}$ consist of the unit cell and hoping matrix of Hamiltonian, respectively. The equation of motion can be rewritten as a transfer matrix form

\begin{equation}\label{eqi3}
\begin{aligned}
\left(\begin{array}{c}
\boldsymbol{C}_{i+1} \\
\boldsymbol{C}_{i}
\end{array}\right)=&
\left(\begin{array}{cc}
\boldsymbol{H}_{i, i+1}^{-1}\left(E \boldsymbol{I}-\boldsymbol{H}_{i}\right) & -\boldsymbol{H}_{i, i+1}^{-1} \boldsymbol{H}_{i, i-1} \\
\boldsymbol{I} & 0
\end{array}\right) \\
&\otimes\left(\begin{array}{c}
\boldsymbol{C}_{i} \\
\boldsymbol{C}_{i-1}
\end{array}\right).
\end{aligned}
\end{equation}

\noindent We suppose the solutions of Eq. (\ref{eqi3}) have Bloch symmetry, $\boldsymbol{C}_{i}=\lambda\boldsymbol{C}_{i-1}$ and $\boldsymbol{C}_{i+1}=\lambda^{2}\boldsymbol{C}_{i-1}$.  Substituting this into Eq. (\ref{eqi3}) results in an eigenvalue problem,

\begin{equation}\label{eqi4}
\begin{aligned}
\left(\begin{array}{cc}
\boldsymbol{H}_{i, i+1}^{-1}\left(E \boldsymbol{I}-\boldsymbol{H}_{i}\right) & -\boldsymbol{H}_{i, i+1}^{-1} \boldsymbol{H}_{i, i-1} \\
\boldsymbol{I} & 0
\end{array}\right)&\left(\begin{array}{c}
\boldsymbol{C}_{i} \\
\boldsymbol{C}_{i-1}
\end{array}\right)=\\
&\lambda\left(\begin{array}{c}
\boldsymbol{C}_{i} \\
\boldsymbol{C}_{i-1}
\end{array}\right).
\end{aligned}
\end{equation}

\noindent The eigenvalue $\lambda$ is related to the wave vector $k$ through $\lambda=\exp (i k a)$. The eigenvalues are denoted by $\lambda_{n}(\pm)$, the corresponding eigenvectors by $\boldsymbol{u}_{n}(\pm)$, where the right-going and left-going modes are labeled as $(+)$ and $(-)$. The Bloch velocities are given by the expression

\begin{equation}\label{eqi5}
v_{n}(\pm)=-\frac{2 a}{\hbar} \operatorname{Im}\left[\lambda_{n}(\pm) \boldsymbol{u}_{n}^{\dagger}(\pm) H_{i, i+1}^{\dagger} \boldsymbol{u}_{n}(\pm)\right].
\end{equation}

\noindent Define

\begin{equation}\label{eqi6}
\boldsymbol{U}({\pm})=\left(\boldsymbol{u}_{\mathbf{1}}(\pm) \cdots \boldsymbol{u}_{3 N_{y}}(\pm)\right)
\end{equation}

\noindent and

\begin{equation}\label{eqi7}
\Lambda(\pm)=\left(\begin{array}{ccc}
\lambda_{1}(\pm) & & \\
& \ddots & \\
& & \lambda_{3 N_{y}}(\pm)
\end{array}\right),
\end{equation}

\noindent where $N_{y}$ is the number of metal atoms in the $y$ direction, then we have

\begin{equation}\label{eqi8}
\boldsymbol{F}({\pm})=\boldsymbol{U}({\pm}) \boldsymbol{\Lambda}({\pm}) \boldsymbol{U}^{-\mathbf{1}}({\pm}).
\end{equation}

The transmission coefficient for the incident mode $m$ with velocity $v_{m}$ and out-going mode $n$ with velocity $v_{n}$ can be obtained as

\begin{equation}\label{eqi9}
\begin{aligned}
t_{m n}=\sqrt{\frac{v_{n}}{v_{m}}}& \left\{ -\boldsymbol{U}^{-1}(+){G}_{N_{x}+1,0}\boldsymbol{H}_{0,-1}\right. \\
& \left.\otimes \left[\boldsymbol{F}^{-1}(+)-\boldsymbol{F}^{-1}(-)\right] \boldsymbol{U}(+)\right\}_{m n},
\end{aligned}
\end{equation}

\noindent and the reflection coefficient for the incident mode $m$ and out-going mode $n$ as

\begin{equation}\label{eqi10}
\begin{aligned}
r_{m n}=\sqrt{\frac{v_{n}}{v_{m}}}& \left( \boldsymbol{U}^{-1}(-)\{-\boldsymbol{G}_{0,0}\boldsymbol{H}_{0,-1}\right.\\
& \left.\otimes \left[\boldsymbol{F}^{-1}(+)-\boldsymbol{F}^{-1}(-)\right] -\boldsymbol{I}\}\boldsymbol{U}(+)\right)_{m n}.
\end{aligned}
\end{equation}

\noindent Here, $N_{x}$ is the number of metal atoms in the $x$ direction and the Green’s-function matrix block $\boldsymbol{G}_{N_{x}+1,0}$ and $\boldsymbol{G}_{0,0}$ can be found using a set of recursive formulas \cite{x46}. Therefore, the valley-dependent transmission and reflection coefficients for the incident valley $A$ and out-going valley $A^{\prime}$ ($A$, $A^{\prime}=K$ or $-K$) can be defined as, respectively,

\begin{equation}\label{eqi11}
T_{A^{\prime}, A}=\sum_{m \in\left\{A^{\prime}\right\}, n \in\{A\}}\left|t_{m n}\right|^{2}
\end{equation}

\noindent and

\begin{equation}\label{eqi12}
R_{A^{\prime}, A}=\sum_{m \in\left\{A^{\prime}\right\}, n \in\{A\}}\left|r_{m n}\right|^{2}.
\end{equation}

\section{BAND STRUCTURE}
\label{BS}

\begin{figure}
\centering
\includegraphics[width=7cm]{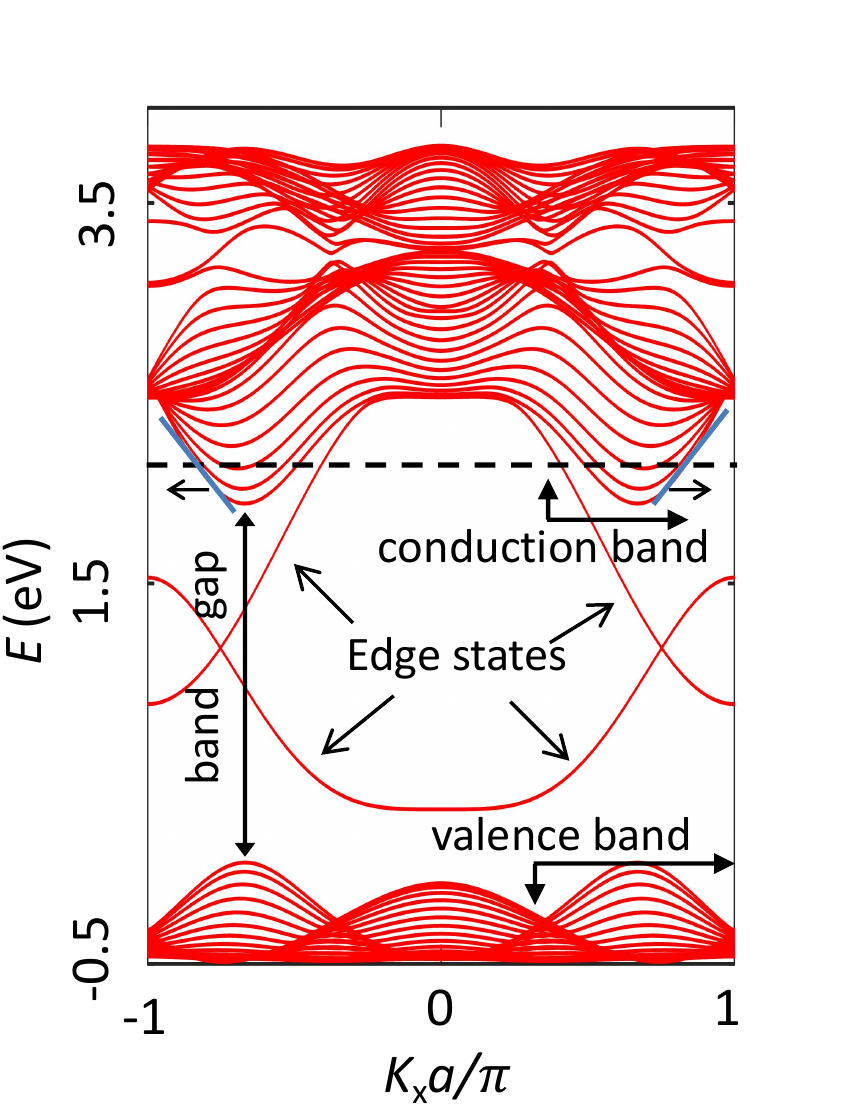}
\caption{\label{fig:a6}The band structure of the nanoribbon without stubs. Black arrows mark valence bands, conduction bands, edge states and band gap. Black dashed line represents the Fermi level, the number of the subbands passing through the Fermi level is the number of modes involved in transport, the $^{\prime}$left-going$^{\prime}$ and $^{\prime}$right-going$^{\prime}$ directions of the group velocity for the states at Fermi energy are determined by the slope of the dispersion. The short blue line shows the tangent line of dispersion, which is represented by $^{\prime} \rightarrow ^{\prime}$ when the slope is greater than 0 and $^{\prime} \leftarrow ^{\prime}$ when the slope is less than 0. $^{\prime} \rightarrow ^{\prime}$ and $^{\prime} \leftarrow ^{\prime}$ represent the direction of movement to the left and right respectively.
}
\end{figure}

In quantum transport theory, there are often multiple modes in system involved in transport.  For each mode $m$, its transmission coefficient $\sum_{n}\left|t_{m n}\right|^{2}$ in Eq. (\ref{eqi11})
is always between 0 and 1. While for transmission of multiple modes, in order to distinguish the contribution of each mode to the transmission, the usual practice is to sum the transmission coefficient of each mode, $\sum_{m,n}\left|t_{m n}\right|^{2}$, as the transmission coefficient for all modes. Therefore, the transmission here is greater than 1. If we assume that each mode in the left lead incident an electron, the transmission coefficient we calculated divided by the number of modes will be between 0 $\sim$ 1.

For the monolayer two-dimensional (2D) $\mathrm{MoS}_{2}$, in both the $K$ and $-K$ valleys, there could be electrons going along all directions in the 2D plane. Here, we study the monolayer $\mathrm{MoS}_{2}$ nanoribbon. For finite width nanoribbons, due to quantum size effects, the two-dimensional band becomes a series of subbands, as shown in Fig.~\ref{fig:a6}. The band gap indicated by double arrows is approximately 1.8 eV, and the black dashed line represents the Fermi level. The number of the subbands passing through the Fermi level is the number of modes involved in transport. When the energy E is between 1.9 and 1.92 eV, only the edge state passes through the Fermi level, which is a single-mode transport. It can be found that in Fig.~\ref{fig:a1}(b), when E is between 1.9 and 1.92 eV, the transmissions between 0 and 1; meanwhile, in  Fig.~\ref{fig:a2}(a), when E is between 1.9 and 1.92 eV, the total transmission is between 0 and 1. In Fig.~\ref{fig:a1}(b), the value of $T_{-K,-K}^\rightarrow$ is greater than 2, indicating that more than two modes are involved in the transport of $-K$ valley.

The band structure in Fig.~\ref{fig:a4}(a) could be understood by folding the one in Fig.~\ref{fig:a6} with the effect of the coupling with the stub. The Brillouin zone folding leads to intersections of bands. At the intersection points of bands of propagating right in the $K$ valley and propagating left in the $-K$ valley, the intervalley backscattering $R_{K,-K}^\leftarrow$ gives rise to the $-K$ valley polarized gap  $\Delta_{+}$. While, the intervalley backscattering $R_{-K,K}^\leftarrow$ results in the  $K$ valley polarized gap $\Delta_{-}$ at the intersection points of bands of propagating right in the $-K$ valley and propagating left in the $K$ valley.

In the quantum confinement direction, we have no way to define the group velocity. According to the definition of group velocity, $v=\frac{1}{\hbar} \frac{\partial E}{\partial k}$, the $^{\prime}$left-going$^{\prime}$ and $^{\prime}$right-going$^{\prime}$ directions of the group velocity for the states at Fermi energy are determined by the slope of the dispersion. $^{\prime} \rightarrow ^{\prime}$ and $^{\prime} \leftarrow ^{\prime}$ represent the direction of movement to the left and right respectively.


\begin{thebibliography}{48}%
\makeatletter
\providecommand \@ifxundefined [1]{%
 \@ifx{#1\undefined}
}%
\providecommand \@ifnum [1]{%
 \ifnum #1\expandafter \@firstoftwo
 \else \expandafter \@secondoftwo
 \fi
}%
\providecommand \@ifx [1]{%
 \ifx #1\expandafter \@firstoftwo
 \else \expandafter \@secondoftwo
 \fi
}%
\providecommand \natexlab [1]{#1}%
\providecommand \enquote  [1]{``#1''}%
\providecommand \bibnamefont  [1]{#1}%
\providecommand \bibfnamefont [1]{#1}%
\providecommand \citenamefont [1]{#1}%
\providecommand \href@noop [0]{\@secondoftwo}%
\providecommand \href [0]{\begingroup \@sanitize@url \@href}%
\providecommand \@href[1]{\@@startlink{#1}\@@href}%
\providecommand \@@href[1]{\endgroup#1\@@endlink}%
\providecommand \@sanitize@url [0]{\catcode `\\12\catcode `\$12\catcode
  `\&12\catcode `\#12\catcode `\^12\catcode `\_12\catcode `\%12\relax}%
\providecommand \@@startlink[1]{}%
\providecommand \@@endlink[0]{}%
\providecommand \url  [0]{\begingroup\@sanitize@url \@url }%
\providecommand \@url [1]{\endgroup\@href {#1}{\urlprefix }}%
\providecommand \urlprefix  [0]{URL }%
\providecommand \Eprint [0]{\href }%
\providecommand \doibase [0]{https://doi.org/}%
\providecommand \selectlanguage [0]{\@gobble}%
\providecommand \bibinfo  [0]{\@secondoftwo}%
\providecommand \bibfield  [0]{\@secondoftwo}%
\providecommand \translation [1]{[#1]}%
\providecommand \BibitemOpen [0]{}%
\providecommand \bibitemStop [0]{}%
\providecommand \bibitemNoStop [0]{.\EOS\space}%
\providecommand \EOS [0]{\spacefactor3000\relax}%
\providecommand \BibitemShut  [1]{\csname bibitem#1\endcsname}%
\let\auto@bib@innerbib\@empty
\bibitem [{\citenamefont {Fano}(1961)}]{x1}%
  \BibitemOpen
  \bibfield  {author} {\bibinfo {author} {\bibfnamefont {U.}~\bibnamefont
  {Fano}},\ }\href@noop {} {\bibfield  {journal} {\bibinfo  {journal} {Phys.
  Rev.}\ }\textbf {\bibinfo {volume} {124}},\ \bibinfo {pages} {1866} (\bibinfo
  {year} {1961})}\BibitemShut {NoStop}%
\bibitem [{\citenamefont {Tekman}\ and\ \citenamefont {Bagwell}(1993)}]{x2}%
  \BibitemOpen
  \bibfield  {author} {\bibinfo {author} {\bibfnamefont {E.}~\bibnamefont
  {Tekman}}\ and\ \bibinfo {author} {\bibfnamefont {P.~F.}\ \bibnamefont
  {Bagwell}},\ }\href@noop {} {\bibfield  {journal} {\bibinfo  {journal} {Phys.
  Rev. B}\ }\textbf {\bibinfo {volume} {48}},\ \bibinfo {pages} {2553}
  (\bibinfo {year} {1993})}\BibitemShut {NoStop}%
\bibitem [{\citenamefont {Kobayashi}\ \emph {et~al.}(2002)\citenamefont
  {Kobayashi}, \citenamefont {Aikawa}, \citenamefont {Katsumoto},\ and\
  \citenamefont {Iye}}]{x3}%
  \BibitemOpen
  \bibfield  {author} {\bibinfo {author} {\bibfnamefont {K.}~\bibnamefont
  {Kobayashi}}, \bibinfo {author} {\bibfnamefont {H.}~\bibnamefont {Aikawa}},
  \bibinfo {author} {\bibfnamefont {S.}~\bibnamefont {Katsumoto}},\ and\
  \bibinfo {author} {\bibfnamefont {Y.}~\bibnamefont {Iye}},\ }\href@noop {}
  {\bibfield  {journal} {\bibinfo  {journal} {Phys. Rev. Lett.}\ }\textbf
  {\bibinfo {volume} {88}},\ \bibinfo {pages} {256806} (\bibinfo {year}
  {2002})}\BibitemShut {NoStop}%
\bibitem [{\citenamefont {Johnson}\ \emph {et~al.}(2004)\citenamefont
  {Johnson}, \citenamefont {Marcus}, \citenamefont {Hanson},\ and\
  \citenamefont {Gossard}}]{x4}%
  \BibitemOpen
  \bibfield  {author} {\bibinfo {author} {\bibfnamefont {A.~C.}\ \bibnamefont
  {Johnson}}, \bibinfo {author} {\bibfnamefont {C.~M.}\ \bibnamefont {Marcus}},
  \bibinfo {author} {\bibfnamefont {M.~P.}\ \bibnamefont {Hanson}},\ and\
  \bibinfo {author} {\bibfnamefont {A.~C.}\ \bibnamefont {Gossard}},\
  }\href@noop {} {\bibfield  {journal} {\bibinfo  {journal} {Phys. Rev. Lett.}\
  }\textbf {\bibinfo {volume} {93}},\ \bibinfo {pages} {106803} (\bibinfo
  {year} {2004})}\BibitemShut {NoStop}%
\bibitem [{\citenamefont {Sánchez}\ and\ \citenamefont {Serra}(2006)}]{x5}%
  \BibitemOpen
  \bibfield  {author} {\bibinfo {author} {\bibfnamefont {D.}~\bibnamefont
  {Sánchez}}\ and\ \bibinfo {author} {\bibfnamefont {L.}~\bibnamefont
  {Serra}},\ }\href@noop {} {\bibfield  {journal} {\bibinfo  {journal} {Phys.
  Rev. B}\ }\textbf {\bibinfo {volume} {74}},\ \bibinfo {pages} {153313}
  (\bibinfo {year} {2006})}\BibitemShut {NoStop}%
\bibitem [{\citenamefont {Recher}\ \emph {et~al.}(2007)\citenamefont {Recher},
  \citenamefont {Trauzettel}, \citenamefont {Rycerz}, \citenamefont {Blanter},
  \citenamefont {Beenakker},\ and\ \citenamefont {Morpurgo}}]{x6}%
  \BibitemOpen
  \bibfield  {author} {\bibinfo {author} {\bibfnamefont {P.}~\bibnamefont
  {Recher}}, \bibinfo {author} {\bibfnamefont {B.}~\bibnamefont {Trauzettel}},
  \bibinfo {author} {\bibfnamefont {A.}~\bibnamefont {Rycerz}}, \bibinfo
  {author} {\bibfnamefont {Y.~M.}\ \bibnamefont {Blanter}}, \bibinfo {author}
  {\bibfnamefont {C.~W.~J.}\ \bibnamefont {Beenakker}},\ and\ \bibinfo {author}
  {\bibfnamefont {A.~F.}\ \bibnamefont {Morpurgo}},\ }\href@noop {} {\bibfield
  {journal} {\bibinfo  {journal} {Phys. Rev. B}\ }\textbf {\bibinfo {volume}
  {76}},\ \bibinfo {pages} {235404} (\bibinfo {year} {2007})}\BibitemShut
  {NoStop}%
\bibitem [{\citenamefont {Zhai}\ and\ \citenamefont {Xu}(2007)}]{x7}%
  \BibitemOpen
  \bibfield  {author} {\bibinfo {author} {\bibfnamefont {F.}~\bibnamefont
  {Zhai}}\ and\ \bibinfo {author} {\bibfnamefont {H.~Q.}\ \bibnamefont {Xu}},\
  }\href@noop {} {\bibfield  {journal} {\bibinfo  {journal} {Phys. Rev. B}\
  }\textbf {\bibinfo {volume} {76}},\ \bibinfo {pages} {035306} (\bibinfo
  {year} {2007})}\BibitemShut {NoStop}%
\bibitem [{\citenamefont {Nowak}\ \emph {et~al.}(2011)\citenamefont {Nowak},
  \citenamefont {Szafran},\ and\ \citenamefont {Peeters}}]{x8}%
  \BibitemOpen
  \bibfield  {author} {\bibinfo {author} {\bibfnamefont {M.~P.}\ \bibnamefont
  {Nowak}}, \bibinfo {author} {\bibfnamefont {B.}~\bibnamefont {Szafran}},\
  and\ \bibinfo {author} {\bibfnamefont {F.~M.}\ \bibnamefont {Peeters}},\
  }\href@noop {} {\bibfield  {journal} {\bibinfo  {journal} {Phys. Rev. B}\
  }\textbf {\bibinfo {volume} {84}},\ \bibinfo {pages} {235319} (\bibinfo
  {year} {2011})}\BibitemShut {NoStop}%
\bibitem [{\citenamefont {Liu}\ \emph {et~al.}(2020)\citenamefont {Liu},
  \citenamefont {Huang}, \citenamefont {Luo},\ and\ \citenamefont {Lai}}]{x9}%
  \BibitemOpen
  \bibfield  {author} {\bibinfo {author} {\bibfnamefont {C.-R.}\ \bibnamefont
  {Liu}}, \bibinfo {author} {\bibfnamefont {L.}~\bibnamefont {Huang}}, \bibinfo
  {author} {\bibfnamefont {H.}~\bibnamefont {Luo}},\ and\ \bibinfo {author}
  {\bibfnamefont {Y.-C.}\ \bibnamefont {Lai}},\ }\href@noop {} {\bibfield
  {journal} {\bibinfo  {journal} {Phys. Rev. Appl.}\ }\textbf {\bibinfo
  {volume} {13}},\ \bibinfo {pages} {034061} (\bibinfo {year}
  {2020})}\BibitemShut {NoStop}%
\bibitem [{\citenamefont {Huang}\ \emph {et~al.}(2021)\citenamefont {Huang},
  \citenamefont {Beyer}, \citenamefont {Puttisong}, \citenamefont {Buyanova},\
  and\ \citenamefont {Chen}}]{x10}%
  \BibitemOpen
  \bibfield  {author} {\bibinfo {author} {\bibfnamefont {Y.~Q.}\ \bibnamefont
  {Huang}}, \bibinfo {author} {\bibfnamefont {J.}~\bibnamefont {Beyer}},
  \bibinfo {author} {\bibfnamefont {Y.}~\bibnamefont {Puttisong}}, \bibinfo
  {author} {\bibfnamefont {I.~A.}\ \bibnamefont {Buyanova}},\ and\ \bibinfo
  {author} {\bibfnamefont {W.~M.}\ \bibnamefont {Chen}},\ }\href@noop {}
  {\bibfield  {journal} {\bibinfo  {journal} {Phys. Rev. Lett.}\ }\textbf
  {\bibinfo {volume} {127}},\ \bibinfo {pages} {127401} (\bibinfo {year}
  {2021})}\BibitemShut {NoStop}%
\bibitem [{\citenamefont {Petrović}\ and\ \citenamefont
  {Peeters}(2015)}]{x11}%
  \BibitemOpen
  \bibfield  {author} {\bibinfo {author} {\bibfnamefont {M.~D.}\ \bibnamefont
  {Petrović}}\ and\ \bibinfo {author} {\bibfnamefont {F.~M.}\ \bibnamefont
  {Peeters}},\ }\href@noop {} {\bibfield  {journal} {\bibinfo  {journal} {Phys.
  Rev. B}\ }\textbf {\bibinfo {volume} {91}},\ \bibinfo {pages} {035444}
  (\bibinfo {year} {2015})}\BibitemShut {NoStop}%
\bibitem [{\citenamefont {Amini}\ \emph {et~al.}(2019)\citenamefont {Amini},
  \citenamefont {Soltani},\ and\ \citenamefont {Sharbafiun}}]{x12}%
  \BibitemOpen
  \bibfield  {author} {\bibinfo {author} {\bibfnamefont {M.}~\bibnamefont
  {Amini}}, \bibinfo {author} {\bibfnamefont {M.}~\bibnamefont {Soltani}},\
  and\ \bibinfo {author} {\bibfnamefont {M.}~\bibnamefont {Sharbafiun}},\
  }\href@noop {} {\bibfield  {journal} {\bibinfo  {journal} {Phys. Rev. B}\
  }\textbf {\bibinfo {volume} {99}},\ \bibinfo {pages} {085403} (\bibinfo
  {year} {2019})}\BibitemShut {NoStop}%
\bibitem [{\citenamefont {Borsoi}\ \emph {et~al.}(2020)\citenamefont {Borsoi},
  \citenamefont {Zuo}, \citenamefont {Gazibegovic}, \citenamefont
  {Op~Het~Veld}, \citenamefont {Bakkers}, \citenamefont {Kouwenhoven},\ and\
  \citenamefont {Heedt}}]{x13}%
  \BibitemOpen
  \bibfield  {author} {\bibinfo {author} {\bibfnamefont {F.}~\bibnamefont
  {Borsoi}}, \bibinfo {author} {\bibfnamefont {K.}~\bibnamefont {Zuo}},
  \bibinfo {author} {\bibfnamefont {S.}~\bibnamefont {Gazibegovic}}, \bibinfo
  {author} {\bibfnamefont {R.~L.~M.}\ \bibnamefont {Op~Het~Veld}}, \bibinfo
  {author} {\bibfnamefont {E.}~\bibnamefont {Bakkers}}, \bibinfo {author}
  {\bibfnamefont {L.~P.}\ \bibnamefont {Kouwenhoven}},\ and\ \bibinfo {author}
  {\bibfnamefont {S.}~\bibnamefont {Heedt}},\ }\href@noop {} {\bibfield
  {journal} {\bibinfo  {journal} {Nat. Commun.}\ }\textbf {\bibinfo {volume}
  {11}},\ \bibinfo {pages} {3666} (\bibinfo {year} {2020})}\BibitemShut
  {NoStop}%
\bibitem [{\citenamefont {Wang}\ \emph
  {et~al.}(2020{\natexlab{a}})\citenamefont {Wang}, \citenamefont {Jin},
  \citenamefont {Wang}, \citenamefont {Bonello}, \citenamefont
  {Djafari-Rouhani},\ and\ \citenamefont {Fleury}}]{x14}%
  \BibitemOpen
  \bibfield  {author} {\bibinfo {author} {\bibfnamefont {W.}~\bibnamefont
  {Wang}}, \bibinfo {author} {\bibfnamefont {Y.}~\bibnamefont {Jin}}, \bibinfo
  {author} {\bibfnamefont {W.}~\bibnamefont {Wang}}, \bibinfo {author}
  {\bibfnamefont {B.}~\bibnamefont {Bonello}}, \bibinfo {author} {\bibfnamefont
  {B.}~\bibnamefont {Djafari-Rouhani}},\ and\ \bibinfo {author} {\bibfnamefont
  {R.}~\bibnamefont {Fleury}},\ }\href@noop {} {\bibfield  {journal} {\bibinfo
  {journal} {Phys. Rev. B}\ }\textbf {\bibinfo {volume} {101}},\ \bibinfo
  {pages} {024101} (\bibinfo {year} {2020}{\natexlab{a}})}\BibitemShut
  {NoStop}%
\bibitem [{\citenamefont {Nikolić}\ and\ \citenamefont {Šordan}(1998)}]{x15}%
  \BibitemOpen
  \bibfield  {author} {\bibinfo {author} {\bibfnamefont {K.}~\bibnamefont
  {Nikolić}}\ and\ \bibinfo {author} {\bibfnamefont {R.}~\bibnamefont
  {Šordan}},\ }\href@noop {} {\bibfield  {journal} {\bibinfo  {journal} {Phys.
  Rev. B}\ }\textbf {\bibinfo {volume} {58}},\ \bibinfo {pages} {9631}
  (\bibinfo {year} {1998})}\BibitemShut {NoStop}%
\bibitem [{\citenamefont {Fan}\ \emph {et~al.}(2014)\citenamefont {Fan},
  \citenamefont {Yu}, \citenamefont {Fan},\ and\ \citenamefont
  {Brongersma}}]{x16}%
  \BibitemOpen
  \bibfield  {author} {\bibinfo {author} {\bibfnamefont {P.}~\bibnamefont
  {Fan}}, \bibinfo {author} {\bibfnamefont {Z.}~\bibnamefont {Yu}}, \bibinfo
  {author} {\bibfnamefont {S.}~\bibnamefont {Fan}},\ and\ \bibinfo {author}
  {\bibfnamefont {M.~L.}\ \bibnamefont {Brongersma}},\ }\href@noop {}
  {\bibfield  {journal} {\bibinfo  {journal} {Nat. Mater.}\ }\textbf {\bibinfo
  {volume} {13}},\ \bibinfo {pages} {471} (\bibinfo {year} {2014})}\BibitemShut
  {NoStop}%
\bibitem [{\citenamefont {Lee}\ \emph {et~al.}(2015)\citenamefont {Lee},
  \citenamefont {Park}, \citenamefont {Han}, \citenamefont {Ee}, \citenamefont
  {Naylor}, \citenamefont {Liu}, \citenamefont {Johnson},\ and\ \citenamefont
  {Agarwal}}]{x17}%
  \BibitemOpen
  \bibfield  {author} {\bibinfo {author} {\bibfnamefont {B.}~\bibnamefont
  {Lee}}, \bibinfo {author} {\bibfnamefont {J.}~\bibnamefont {Park}}, \bibinfo
  {author} {\bibfnamefont {G.~H.}\ \bibnamefont {Han}}, \bibinfo {author}
  {\bibfnamefont {H.~S.}\ \bibnamefont {Ee}}, \bibinfo {author} {\bibfnamefont
  {C.~H.}\ \bibnamefont {Naylor}}, \bibinfo {author} {\bibfnamefont
  {W.}~\bibnamefont {Liu}}, \bibinfo {author} {\bibfnamefont {A.~T.}\
  \bibnamefont {Johnson}},\ and\ \bibinfo {author} {\bibfnamefont
  {R.}~\bibnamefont {Agarwal}},\ }\href@noop {} {\bibfield  {journal} {\bibinfo
   {journal} {Nano Lett.}\ }\textbf {\bibinfo {volume} {15}},\ \bibinfo {pages}
  {3646} (\bibinfo {year} {2015})}\BibitemShut {NoStop}%
\bibitem [{\citenamefont {Myoung}\ \emph {et~al.}(2019)\citenamefont {Myoung},
  \citenamefont {Ryu}, \citenamefont {Park}, \citenamefont {Lee},\ and\
  \citenamefont {Woo}}]{x18}%
  \BibitemOpen
  \bibfield  {author} {\bibinfo {author} {\bibfnamefont {N.}~\bibnamefont
  {Myoung}}, \bibinfo {author} {\bibfnamefont {J.-W.}\ \bibnamefont {Ryu}},
  \bibinfo {author} {\bibfnamefont {H.~C.}\ \bibnamefont {Park}}, \bibinfo
  {author} {\bibfnamefont {S.~J.}\ \bibnamefont {Lee}},\ and\ \bibinfo {author}
  {\bibfnamefont {S.}~\bibnamefont {Woo}},\ }\href@noop {} {\bibfield
  {journal} {\bibinfo  {journal} {Phys. Rev. B}\ }\textbf {\bibinfo {volume}
  {100}},\ \bibinfo {pages} {045427} (\bibinfo {year} {2019})}\BibitemShut
  {NoStop}%
\bibitem [{\citenamefont {Wang}\ \emph {et~al.}(2018)\citenamefont {Wang},
  \citenamefont {Krasnok}, \citenamefont {Zhang}, \citenamefont {Scarabelli},
  \citenamefont {Liu}, \citenamefont {Wu}, \citenamefont {Liz-Marzán},
  \citenamefont {Terrones}, \citenamefont {Alù},\ and\ \citenamefont
  {Zheng}}]{x19}%
  \BibitemOpen
  \bibfield  {author} {\bibinfo {author} {\bibfnamefont {M.}~\bibnamefont
  {Wang}}, \bibinfo {author} {\bibfnamefont {A.}~\bibnamefont {Krasnok}},
  \bibinfo {author} {\bibfnamefont {T.}~\bibnamefont {Zhang}}, \bibinfo
  {author} {\bibfnamefont {L.}~\bibnamefont {Scarabelli}}, \bibinfo {author}
  {\bibfnamefont {H.}~\bibnamefont {Liu}}, \bibinfo {author} {\bibfnamefont
  {Z.}~\bibnamefont {Wu}}, \bibinfo {author} {\bibfnamefont {L.~M.}\
  \bibnamefont {Liz-Marzán}}, \bibinfo {author} {\bibfnamefont
  {M.}~\bibnamefont {Terrones}}, \bibinfo {author} {\bibfnamefont
  {A.}~\bibnamefont {Alù}},\ and\ \bibinfo {author} {\bibfnamefont
  {Y.}~\bibnamefont {Zheng}},\ }\href@noop {} {\bibfield  {journal} {\bibinfo
  {journal} {Adv. Mater.}\ }\textbf {\bibinfo {volume} {30}},\ \bibinfo {pages}
  {1705779} (\bibinfo {year} {2018})}\BibitemShut {NoStop}%
\bibitem [{\citenamefont {Rycerz}\ \emph {et~al.}(2007)\citenamefont {Rycerz},
  \citenamefont {Tworzydło},\ and\ \citenamefont {Beenakker}}]{x20}%
  \BibitemOpen
  \bibfield  {author} {\bibinfo {author} {\bibfnamefont {A.}~\bibnamefont
  {Rycerz}}, \bibinfo {author} {\bibfnamefont {J.}~\bibnamefont {Tworzydło}},\
  and\ \bibinfo {author} {\bibfnamefont {C.~W.~J.}\ \bibnamefont {Beenakker}},\
  }\href@noop {} {\bibfield  {journal} {\bibinfo  {journal} {Nat. Phys.}\
  }\textbf {\bibinfo {volume} {3}},\ \bibinfo {pages} {172} (\bibinfo {year}
  {2007})}\BibitemShut {NoStop}%
\bibitem [{\citenamefont {Singh}\ and\ \citenamefont
  {Schwingenschlögl}(2017)}]{x21}%
  \BibitemOpen
  \bibfield  {author} {\bibinfo {author} {\bibfnamefont {N.}~\bibnamefont
  {Singh}}\ and\ \bibinfo {author} {\bibfnamefont {U.}~\bibnamefont
  {Schwingenschlögl}},\ }\href@noop {} {\bibfield  {journal} {\bibinfo
  {journal} {Adv. Mater.}\ }\textbf {\bibinfo {volume} {29}},\ \bibinfo {pages}
  {1600970} (\bibinfo {year} {2017})}\BibitemShut {NoStop}%
\bibitem [{\citenamefont {Gunlycke}\ and\ \citenamefont {White}(2011)}]{x22}%
  \BibitemOpen
  \bibfield  {author} {\bibinfo {author} {\bibfnamefont {D.}~\bibnamefont
  {Gunlycke}}\ and\ \bibinfo {author} {\bibfnamefont {C.~T.}\ \bibnamefont
  {White}},\ }\href@noop {} {\bibfield  {journal} {\bibinfo  {journal} {Phys.
  Rev. Lett.}\ }\textbf {\bibinfo {volume} {106}},\ \bibinfo {pages} {136806}
  (\bibinfo {year} {2011})}\BibitemShut {NoStop}%
\bibitem [{\citenamefont {Wang}\ \emph
  {et~al.}(2020{\natexlab{b}})\citenamefont {Wang}, \citenamefont {Deng},
  \citenamefont {Wei}, \citenamefont {Wan}, \citenamefont {Liu}, \citenamefont
  {Lu}, \citenamefont {Li}, \citenamefont {Bi}, \citenamefont {Zhang},
  \citenamefont {Lu}, \citenamefont {Chen}, \citenamefont {Zhou}, \citenamefont
  {Zhang}, \citenamefont {Cheng}, \citenamefont {Zhao}, \citenamefont {Ye},
  \citenamefont {Huang}, \citenamefont {Pennycook}, \citenamefont {Loh},\ and\
  \citenamefont {Peng}}]{x23}%
  \BibitemOpen
  \bibfield  {author} {\bibinfo {author} {\bibfnamefont {Y.}~\bibnamefont
  {Wang}}, \bibinfo {author} {\bibfnamefont {L.}~\bibnamefont {Deng}}, \bibinfo
  {author} {\bibfnamefont {Q.}~\bibnamefont {Wei}}, \bibinfo {author}
  {\bibfnamefont {Y.}~\bibnamefont {Wan}}, \bibinfo {author} {\bibfnamefont
  {Z.}~\bibnamefont {Liu}}, \bibinfo {author} {\bibfnamefont {X.}~\bibnamefont
  {Lu}}, \bibinfo {author} {\bibfnamefont {Y.}~\bibnamefont {Li}}, \bibinfo
  {author} {\bibfnamefont {L.}~\bibnamefont {Bi}}, \bibinfo {author}
  {\bibfnamefont {L.}~\bibnamefont {Zhang}}, \bibinfo {author} {\bibfnamefont
  {H.}~\bibnamefont {Lu}}, \bibinfo {author} {\bibfnamefont {H.}~\bibnamefont
  {Chen}}, \bibinfo {author} {\bibfnamefont {P.}~\bibnamefont {Zhou}}, \bibinfo
  {author} {\bibfnamefont {L.}~\bibnamefont {Zhang}}, \bibinfo {author}
  {\bibfnamefont {Y.}~\bibnamefont {Cheng}}, \bibinfo {author} {\bibfnamefont
  {X.}~\bibnamefont {Zhao}}, \bibinfo {author} {\bibfnamefont {Y.}~\bibnamefont
  {Ye}}, \bibinfo {author} {\bibfnamefont {W.}~\bibnamefont {Huang}}, \bibinfo
  {author} {\bibfnamefont {S.~J.}\ \bibnamefont {Pennycook}}, \bibinfo {author}
  {\bibfnamefont {K.~P.}\ \bibnamefont {Loh}},\ and\ \bibinfo {author}
  {\bibfnamefont {B.}~\bibnamefont {Peng}},\ }\href@noop {} {\bibfield
  {journal} {\bibinfo  {journal} {Nano Lett.}\ }\textbf {\bibinfo {volume}
  {20}},\ \bibinfo {pages} {2129} (\bibinfo {year}
  {2020}{\natexlab{b}})}\BibitemShut {NoStop}%
\bibitem [{\citenamefont {Low}\ and\ \citenamefont {Guinea}(2010)}]{x24}%
  \BibitemOpen
  \bibfield  {author} {\bibinfo {author} {\bibfnamefont {T.}~\bibnamefont
  {Low}}\ and\ \bibinfo {author} {\bibfnamefont {F.}~\bibnamefont {Guinea}},\
  }\href@noop {} {\bibfield  {journal} {\bibinfo  {journal} {Nano Lett.}\
  }\textbf {\bibinfo {volume} {10}},\ \bibinfo {pages} {3551} (\bibinfo {year}
  {2010})}\BibitemShut {NoStop}%
\bibitem [{\citenamefont {Myoung}\ \emph {et~al.}(2020)\citenamefont {Myoung},
  \citenamefont {Choi},\ and\ \citenamefont {Park}}]{x25}%
  \BibitemOpen
  \bibfield  {author} {\bibinfo {author} {\bibfnamefont {N.}~\bibnamefont
  {Myoung}}, \bibinfo {author} {\bibfnamefont {H.}~\bibnamefont {Choi}},\ and\
  \bibinfo {author} {\bibfnamefont {H.~C.}\ \bibnamefont {Park}},\ }\href@noop
  {} {\bibfield  {journal} {\bibinfo  {journal} {Carbon}\ }\textbf {\bibinfo
  {volume} {157}},\ \bibinfo {pages} {578} (\bibinfo {year}
  {2020})}\BibitemShut {NoStop}%
\bibitem [{\citenamefont {Wu}\ \emph {et~al.}(2011)\citenamefont {Wu},
  \citenamefont {Zhai}, \citenamefont {Peeters}, \citenamefont {Xu},\ and\
  \citenamefont {Chang}}]{x26}%
  \BibitemOpen
  \bibfield  {author} {\bibinfo {author} {\bibfnamefont {Z.}~\bibnamefont
  {Wu}}, \bibinfo {author} {\bibfnamefont {F.}~\bibnamefont {Zhai}}, \bibinfo
  {author} {\bibfnamefont {F.~M.}\ \bibnamefont {Peeters}}, \bibinfo {author}
  {\bibfnamefont {H.~Q.}\ \bibnamefont {Xu}},\ and\ \bibinfo {author}
  {\bibfnamefont {K.}~\bibnamefont {Chang}},\ }\href@noop {} {\bibfield
  {journal} {\bibinfo  {journal} {Phys. Rev. Lett.}\ }\textbf {\bibinfo
  {volume} {106}},\ \bibinfo {pages} {176802} (\bibinfo {year}
  {2011})}\BibitemShut {NoStop}%
\bibitem [{\citenamefont {Georgi}\ \emph {et~al.}(2017)\citenamefont {Georgi},
  \citenamefont {Nemes-Incze}, \citenamefont {Carrillo-Bastos}, \citenamefont
  {Faria}, \citenamefont {Viola~Kusminskiy}, \citenamefont {Zhai},
  \citenamefont {Schneider}, \citenamefont {Subramaniam}, \citenamefont
  {Mashoff}, \citenamefont {Freitag}, \citenamefont {Liebmann}, \citenamefont
  {Pratzer}, \citenamefont {Wirtz}, \citenamefont {Woods}, \citenamefont
  {Gorbachev}, \citenamefont {Cao}, \citenamefont {Novoselov}, \citenamefont
  {Sandler},\ and\ \citenamefont {Morgenstern}}]{x27}%
  \BibitemOpen
  \bibfield  {author} {\bibinfo {author} {\bibfnamefont {A.}~\bibnamefont
  {Georgi}}, \bibinfo {author} {\bibfnamefont {P.}~\bibnamefont {Nemes-Incze}},
  \bibinfo {author} {\bibfnamefont {R.}~\bibnamefont {Carrillo-Bastos}},
  \bibinfo {author} {\bibfnamefont {D.}~\bibnamefont {Faria}}, \bibinfo
  {author} {\bibfnamefont {S.}~\bibnamefont {Viola~Kusminskiy}}, \bibinfo
  {author} {\bibfnamefont {D.}~\bibnamefont {Zhai}}, \bibinfo {author}
  {\bibfnamefont {M.}~\bibnamefont {Schneider}}, \bibinfo {author}
  {\bibfnamefont {D.}~\bibnamefont {Subramaniam}}, \bibinfo {author}
  {\bibfnamefont {T.}~\bibnamefont {Mashoff}}, \bibinfo {author} {\bibfnamefont
  {N.~M.}\ \bibnamefont {Freitag}}, \bibinfo {author} {\bibfnamefont
  {M.}~\bibnamefont {Liebmann}}, \bibinfo {author} {\bibfnamefont
  {M.}~\bibnamefont {Pratzer}}, \bibinfo {author} {\bibfnamefont
  {L.}~\bibnamefont {Wirtz}}, \bibinfo {author} {\bibfnamefont {C.~R.}\
  \bibnamefont {Woods}}, \bibinfo {author} {\bibfnamefont {R.~V.}\ \bibnamefont
  {Gorbachev}}, \bibinfo {author} {\bibfnamefont {Y.}~\bibnamefont {Cao}},
  \bibinfo {author} {\bibfnamefont {K.~S.}\ \bibnamefont {Novoselov}}, \bibinfo
  {author} {\bibfnamefont {N.}~\bibnamefont {Sandler}},\ and\ \bibinfo {author}
  {\bibfnamefont {M.}~\bibnamefont {Morgenstern}},\ }\href@noop {} {\bibfield
  {journal} {\bibinfo  {journal} {Nano Lett.}\ }\textbf {\bibinfo {volume}
  {17}},\ \bibinfo {pages} {2240} (\bibinfo {year} {2017})}\BibitemShut
  {NoStop}%
\bibitem [{\citenamefont {Li}\ \emph {et~al.}(2020{\natexlab{a}})\citenamefont
  {Li}, \citenamefont {Su}, \citenamefont {Ren},\ and\ \citenamefont
  {He}}]{x28}%
  \BibitemOpen
  \bibfield  {author} {\bibinfo {author} {\bibfnamefont {S.~Y.}\ \bibnamefont
  {Li}}, \bibinfo {author} {\bibfnamefont {Y.}~\bibnamefont {Su}}, \bibinfo
  {author} {\bibfnamefont {Y.~N.}\ \bibnamefont {Ren}},\ and\ \bibinfo {author}
  {\bibfnamefont {L.}~\bibnamefont {He}},\ }\href@noop {} {\bibfield  {journal}
  {\bibinfo  {journal} {Phys. Rev. Lett.}\ }\textbf {\bibinfo {volume} {124}},\
  \bibinfo {pages} {106802} (\bibinfo {year} {2020}{\natexlab{a}})}\BibitemShut
  {NoStop}%
\bibitem [{\citenamefont {An}\ \emph {et~al.}(2017)\citenamefont {An},
  \citenamefont {Xiao}, \citenamefont {Tu}, \citenamefont {Yu}, \citenamefont
  {Fal’ko},\ and\ \citenamefont {Yao}}]{x29}%
  \BibitemOpen
  \bibfield  {author} {\bibinfo {author} {\bibfnamefont {X.-T.}\ \bibnamefont
  {An}}, \bibinfo {author} {\bibfnamefont {J.}~\bibnamefont {Xiao}}, \bibinfo
  {author} {\bibfnamefont {M.~W.~Y.}\ \bibnamefont {Tu}}, \bibinfo {author}
  {\bibfnamefont {H.}~\bibnamefont {Yu}}, \bibinfo {author} {\bibfnamefont
  {V.~I.}\ \bibnamefont {Fal’ko}},\ and\ \bibinfo {author} {\bibfnamefont
  {W.}~\bibnamefont {Yao}},\ }\href@noop {} {\bibfield  {journal} {\bibinfo
  {journal} {Phys. Rev. Lett.}\ }\textbf {\bibinfo {volume} {118}},\ \bibinfo
  {pages} {096602} (\bibinfo {year} {2017})}\BibitemShut {NoStop}%
\bibitem [{\citenamefont {An}\ and\ \citenamefont {Yao}(2020)}]{x30}%
  \BibitemOpen
  \bibfield  {author} {\bibinfo {author} {\bibfnamefont {X.-T.}\ \bibnamefont
  {An}}\ and\ \bibinfo {author} {\bibfnamefont {W.}~\bibnamefont {Yao}},\
  }\href@noop {} {\bibfield  {journal} {\bibinfo  {journal} {Phys. Rev. Appl.}\
  }\textbf {\bibinfo {volume} {14}},\ \bibinfo {pages} {014039} (\bibinfo
  {year} {2020})}\BibitemShut {NoStop}%
\bibitem [{\citenamefont {Yu}\ \emph {et~al.}(2014)\citenamefont {Yu},
  \citenamefont {Wu}, \citenamefont {Liu}, \citenamefont {Xu},\ and\
  \citenamefont {Yao}}]{x31}%
  \BibitemOpen
  \bibfield  {author} {\bibinfo {author} {\bibfnamefont {H.}~\bibnamefont
  {Yu}}, \bibinfo {author} {\bibfnamefont {Y.}~\bibnamefont {Wu}}, \bibinfo
  {author} {\bibfnamefont {G.-B.}\ \bibnamefont {Liu}}, \bibinfo {author}
  {\bibfnamefont {X.}~\bibnamefont {Xu}},\ and\ \bibinfo {author}
  {\bibfnamefont {W.}~\bibnamefont {Yao}},\ }\href@noop {} {\bibfield
  {journal} {\bibinfo  {journal} {Phys. Rev. Lett.}\ }\textbf {\bibinfo
  {volume} {113}},\ \bibinfo {pages} {156603} (\bibinfo {year}
  {2014})}\BibitemShut {NoStop}%
\bibitem [{\citenamefont {Sui}\ \emph {et~al.}(2015)\citenamefont {Sui},
  \citenamefont {Chen}, \citenamefont {Ma}, \citenamefont {Shan}, \citenamefont
  {Tian}, \citenamefont {Watanabe}, \citenamefont {Taniguchi}, \citenamefont
  {Jin}, \citenamefont {Yao}, \citenamefont {Xiao},\ and\ \citenamefont
  {Zhang}}]{x32}%
  \BibitemOpen
  \bibfield  {author} {\bibinfo {author} {\bibfnamefont {M.}~\bibnamefont
  {Sui}}, \bibinfo {author} {\bibfnamefont {G.}~\bibnamefont {Chen}}, \bibinfo
  {author} {\bibfnamefont {L.}~\bibnamefont {Ma}}, \bibinfo {author}
  {\bibfnamefont {W.-Y.}\ \bibnamefont {Shan}}, \bibinfo {author}
  {\bibfnamefont {D.}~\bibnamefont {Tian}}, \bibinfo {author} {\bibfnamefont
  {K.}~\bibnamefont {Watanabe}}, \bibinfo {author} {\bibfnamefont
  {T.}~\bibnamefont {Taniguchi}}, \bibinfo {author} {\bibfnamefont
  {X.}~\bibnamefont {Jin}}, \bibinfo {author} {\bibfnamefont {W.}~\bibnamefont
  {Yao}}, \bibinfo {author} {\bibfnamefont {D.}~\bibnamefont {Xiao}},\ and\
  \bibinfo {author} {\bibfnamefont {Y.}~\bibnamefont {Zhang}},\ }\href@noop {}
  {\bibfield  {journal} {\bibinfo  {journal} {Nat. Phys.}\ }\textbf {\bibinfo
  {volume} {11}},\ \bibinfo {pages} {1027} (\bibinfo {year}
  {2015})}\BibitemShut {NoStop}%
\bibitem [{\citenamefont {Mak}\ \emph {et~al.}(2014)\citenamefont {Mak},
  \citenamefont {McGill}, \citenamefont {Park},\ and\ \citenamefont
  {McEuen}}]{x33}%
  \BibitemOpen
  \bibfield  {author} {\bibinfo {author} {\bibfnamefont {K.~F.}\ \bibnamefont
  {Mak}}, \bibinfo {author} {\bibfnamefont {K.~L.}\ \bibnamefont {McGill}},
  \bibinfo {author} {\bibfnamefont {J.}~\bibnamefont {Park}},\ and\ \bibinfo
  {author} {\bibfnamefont {P.~L.}\ \bibnamefont {McEuen}},\ }\href@noop {}
  {\bibfield  {journal} {\bibinfo  {journal} {Science}\ }\textbf {\bibinfo
  {volume} {344}},\ \bibinfo {pages} {1489} (\bibinfo {year}
  {2014})}\BibitemShut {NoStop}%
\bibitem [{\citenamefont {Komatsu}\ \emph {et~al.}(2018)\citenamefont
  {Komatsu}, \citenamefont {Morita}, \citenamefont {Watanabe}, \citenamefont
  {Tsuya}, \citenamefont {Watanabe}, \citenamefont {Taniguchi},\ and\
  \citenamefont {Moriyama}}]{x34}%
  \BibitemOpen
  \bibfield  {author} {\bibinfo {author} {\bibfnamefont {K.}~\bibnamefont
  {Komatsu}}, \bibinfo {author} {\bibfnamefont {Y.}~\bibnamefont {Morita}},
  \bibinfo {author} {\bibfnamefont {E.}~\bibnamefont {Watanabe}}, \bibinfo
  {author} {\bibfnamefont {D.}~\bibnamefont {Tsuya}}, \bibinfo {author}
  {\bibfnamefont {K.}~\bibnamefont {Watanabe}}, \bibinfo {author}
  {\bibfnamefont {T.}~\bibnamefont {Taniguchi}},\ and\ \bibinfo {author}
  {\bibfnamefont {S.}~\bibnamefont {Moriyama}},\ }\href@noop {} {\bibfield
  {journal} {\bibinfo  {journal} {Sci. Adv.}\ }\textbf {\bibinfo {volume}
  {4}},\ \bibinfo {pages} {eaaq0194} (\bibinfo {year} {2018})}\BibitemShut
  {NoStop}%
\bibitem [{\citenamefont {Back}\ \emph {et~al.}(2017)\citenamefont {Back},
  \citenamefont {Sidler}, \citenamefont {Cotlet}, \citenamefont {Srivastava},
  \citenamefont {Takemura}, \citenamefont {Kroner},\ and\ \citenamefont
  {Imamoglu}}]{x35}%
  \BibitemOpen
  \bibfield  {author} {\bibinfo {author} {\bibfnamefont {P.}~\bibnamefont
  {Back}}, \bibinfo {author} {\bibfnamefont {M.}~\bibnamefont {Sidler}},
  \bibinfo {author} {\bibfnamefont {O.}~\bibnamefont {Cotlet}}, \bibinfo
  {author} {\bibfnamefont {A.}~\bibnamefont {Srivastava}}, \bibinfo {author}
  {\bibfnamefont {N.}~\bibnamefont {Takemura}}, \bibinfo {author}
  {\bibfnamefont {M.}~\bibnamefont {Kroner}},\ and\ \bibinfo {author}
  {\bibfnamefont {A.}~\bibnamefont {Imamoglu}},\ }\href@noop {} {\bibfield
  {journal} {\bibinfo  {journal} {Phys. Rev. Lett.}\ }\textbf {\bibinfo
  {volume} {118}},\ \bibinfo {pages} {237404} (\bibinfo {year}
  {2017})}\BibitemShut {NoStop}%
\bibitem [{\citenamefont {Ye}\ \emph {et~al.}(2016)\citenamefont {Ye},
  \citenamefont {Xiao}, \citenamefont {Wang}, \citenamefont {Ye}, \citenamefont
  {Zhu}, \citenamefont {Zhao}, \citenamefont {Wang}, \citenamefont {Zhao},
  \citenamefont {Yin},\ and\ \citenamefont {Zhang}}]{x36}%
  \BibitemOpen
  \bibfield  {author} {\bibinfo {author} {\bibfnamefont {Y.}~\bibnamefont
  {Ye}}, \bibinfo {author} {\bibfnamefont {J.}~\bibnamefont {Xiao}}, \bibinfo
  {author} {\bibfnamefont {H.}~\bibnamefont {Wang}}, \bibinfo {author}
  {\bibfnamefont {Z.}~\bibnamefont {Ye}}, \bibinfo {author} {\bibfnamefont
  {H.}~\bibnamefont {Zhu}}, \bibinfo {author} {\bibfnamefont {M.}~\bibnamefont
  {Zhao}}, \bibinfo {author} {\bibfnamefont {Y.}~\bibnamefont {Wang}}, \bibinfo
  {author} {\bibfnamefont {J.}~\bibnamefont {Zhao}}, \bibinfo {author}
  {\bibfnamefont {X.}~\bibnamefont {Yin}},\ and\ \bibinfo {author}
  {\bibfnamefont {X.}~\bibnamefont {Zhang}},\ }\href@noop {} {\bibfield
  {journal} {\bibinfo  {journal} {Nat. Nanotech.}\ }\textbf {\bibinfo {volume}
  {11}},\ \bibinfo {pages} {598} (\bibinfo {year} {2016})}\BibitemShut
  {NoStop}%
\bibitem [{\citenamefont {Zhong}\ \emph {et~al.}(2017)\citenamefont {Zhong},
  \citenamefont {Seyler}, \citenamefont {Linpeng}, \citenamefont {Cheng},
  \citenamefont {Sivadas}, \citenamefont {Huang}, \citenamefont {Schmidgall},
  \citenamefont {Taniguchi}, \citenamefont {Watanabe}, \citenamefont {McGuire},
  \citenamefont {Yao}, \citenamefont {Xiao}, \citenamefont {Fu},\ and\
  \citenamefont {Xu}}]{x37}%
  \BibitemOpen
  \bibfield  {author} {\bibinfo {author} {\bibfnamefont {D.}~\bibnamefont
  {Zhong}}, \bibinfo {author} {\bibfnamefont {K.~L.}\ \bibnamefont {Seyler}},
  \bibinfo {author} {\bibfnamefont {X.}~\bibnamefont {Linpeng}}, \bibinfo
  {author} {\bibfnamefont {R.}~\bibnamefont {Cheng}}, \bibinfo {author}
  {\bibfnamefont {N.}~\bibnamefont {Sivadas}}, \bibinfo {author} {\bibfnamefont
  {B.}~\bibnamefont {Huang}}, \bibinfo {author} {\bibfnamefont
  {E.}~\bibnamefont {Schmidgall}}, \bibinfo {author} {\bibfnamefont
  {T.}~\bibnamefont {Taniguchi}}, \bibinfo {author} {\bibfnamefont
  {K.}~\bibnamefont {Watanabe}}, \bibinfo {author} {\bibfnamefont {M.~A.}\
  \bibnamefont {McGuire}}, \bibinfo {author} {\bibfnamefont {W.}~\bibnamefont
  {Yao}}, \bibinfo {author} {\bibfnamefont {D.}~\bibnamefont {Xiao}}, \bibinfo
  {author} {\bibfnamefont {K.-M.~C.}\ \bibnamefont {Fu}},\ and\ \bibinfo
  {author} {\bibfnamefont {X.}~\bibnamefont {Xu}},\ }\href@noop {} {\bibfield
  {journal} {\bibinfo  {journal} {Sci. Adv.}\ }\textbf {\bibinfo {volume}
  {3}},\ \bibinfo {pages} {e1603113} (\bibinfo {year} {2017})}\BibitemShut
  {NoStop}%
\bibitem [{\citenamefont {Nagler}\ \emph {et~al.}(2017)\citenamefont {Nagler},
  \citenamefont {Ballottin}, \citenamefont {Mitioglu}, \citenamefont
  {Mooshammer}, \citenamefont {Paradiso}, \citenamefont {Strunk}, \citenamefont
  {Huber}, \citenamefont {Chernikov}, \citenamefont {Christianen},
  \citenamefont {Schuller},\ and\ \citenamefont {Korn}}]{x38}%
  \BibitemOpen
  \bibfield  {author} {\bibinfo {author} {\bibfnamefont {P.}~\bibnamefont
  {Nagler}}, \bibinfo {author} {\bibfnamefont {M.~V.}\ \bibnamefont
  {Ballottin}}, \bibinfo {author} {\bibfnamefont {A.~A.}\ \bibnamefont
  {Mitioglu}}, \bibinfo {author} {\bibfnamefont {F.}~\bibnamefont
  {Mooshammer}}, \bibinfo {author} {\bibfnamefont {N.}~\bibnamefont
  {Paradiso}}, \bibinfo {author} {\bibfnamefont {C.}~\bibnamefont {Strunk}},
  \bibinfo {author} {\bibfnamefont {R.}~\bibnamefont {Huber}}, \bibinfo
  {author} {\bibfnamefont {A.}~\bibnamefont {Chernikov}}, \bibinfo {author}
  {\bibfnamefont {P.~C.~M.}\ \bibnamefont {Christianen}}, \bibinfo {author}
  {\bibfnamefont {C.}~\bibnamefont {Schuller}},\ and\ \bibinfo {author}
  {\bibfnamefont {T.}~\bibnamefont {Korn}},\ }\href@noop {} {\bibfield
  {journal} {\bibinfo  {journal} {Nat. Commun.}\ }\textbf {\bibinfo {volume}
  {8}},\ \bibinfo {pages} {1551} (\bibinfo {year} {2017})}\BibitemShut
  {NoStop}%
\bibitem [{\citenamefont {Li}\ \emph {et~al.}(2020{\natexlab{b}})\citenamefont
  {Li}, \citenamefont {Zhao}, \citenamefont {Deng}, \citenamefont {Shi},
  \citenamefont {Liu}, \citenamefont {Wei}, \citenamefont {Zhang},
  \citenamefont {Cheng}, \citenamefont {Zhang}, \citenamefont {Lu},
  \citenamefont {Gao}, \citenamefont {Huang}, \citenamefont {Qiu},
  \citenamefont {Xiang}, \citenamefont {Pennycook}, \citenamefont {Xiong},
  \citenamefont {Loh},\ and\ \citenamefont {Peng}}]{x39}%
  \BibitemOpen
  \bibfield  {author} {\bibinfo {author} {\bibfnamefont {Q.}~\bibnamefont
  {Li}}, \bibinfo {author} {\bibfnamefont {X.}~\bibnamefont {Zhao}}, \bibinfo
  {author} {\bibfnamefont {L.}~\bibnamefont {Deng}}, \bibinfo {author}
  {\bibfnamefont {Z.}~\bibnamefont {Shi}}, \bibinfo {author} {\bibfnamefont
  {S.}~\bibnamefont {Liu}}, \bibinfo {author} {\bibfnamefont {Q.}~\bibnamefont
  {Wei}}, \bibinfo {author} {\bibfnamefont {L.}~\bibnamefont {Zhang}}, \bibinfo
  {author} {\bibfnamefont {Y.}~\bibnamefont {Cheng}}, \bibinfo {author}
  {\bibfnamefont {L.}~\bibnamefont {Zhang}}, \bibinfo {author} {\bibfnamefont
  {H.}~\bibnamefont {Lu}}, \bibinfo {author} {\bibfnamefont {W.}~\bibnamefont
  {Gao}}, \bibinfo {author} {\bibfnamefont {W.}~\bibnamefont {Huang}}, \bibinfo
  {author} {\bibfnamefont {C.~W.}\ \bibnamefont {Qiu}}, \bibinfo {author}
  {\bibfnamefont {G.}~\bibnamefont {Xiang}}, \bibinfo {author} {\bibfnamefont
  {S.~J.}\ \bibnamefont {Pennycook}}, \bibinfo {author} {\bibfnamefont
  {Q.}~\bibnamefont {Xiong}}, \bibinfo {author} {\bibfnamefont {K.~P.}\
  \bibnamefont {Loh}},\ and\ \bibinfo {author} {\bibfnamefont {B.}~\bibnamefont
  {Peng}},\ }\href@noop {} {\bibfield  {journal} {\bibinfo  {journal} {ACS
  Nano}\ }\textbf {\bibinfo {volume} {14}},\ \bibinfo {pages} {4636} (\bibinfo
  {year} {2020}{\natexlab{b}})}\BibitemShut {NoStop}%
\bibitem [{\citenamefont {Li}\ \emph {et~al.}(2020{\natexlab{c}})\citenamefont
  {Li}, \citenamefont {Jiang}, \citenamefont {Wang}, \citenamefont {Watanabe},
  \citenamefont {Taniguchi}, \citenamefont {Shan},\ and\ \citenamefont
  {Mak}}]{x40}%
  \BibitemOpen
  \bibfield  {author} {\bibinfo {author} {\bibfnamefont {L.}~\bibnamefont
  {Li}}, \bibinfo {author} {\bibfnamefont {S.}~\bibnamefont {Jiang}}, \bibinfo
  {author} {\bibfnamefont {Z.}~\bibnamefont {Wang}}, \bibinfo {author}
  {\bibfnamefont {K.}~\bibnamefont {Watanabe}}, \bibinfo {author}
  {\bibfnamefont {T.}~\bibnamefont {Taniguchi}}, \bibinfo {author}
  {\bibfnamefont {J.}~\bibnamefont {Shan}},\ and\ \bibinfo {author}
  {\bibfnamefont {K.~F.}\ \bibnamefont {Mak}},\ }\href@noop {} {\bibfield
  {journal} {\bibinfo  {journal} {Phys. Rev. Mater.}\ }\textbf {\bibinfo
  {volume} {4}},\ \bibinfo {pages} {104005} (\bibinfo {year}
  {2020}{\natexlab{c}})}\BibitemShut {NoStop}%
\bibitem [{\citenamefont {Zhang}\ \emph {et~al.}(2019)\citenamefont {Zhang},
  \citenamefont {Nie}, \citenamefont {Sanvito},\ and\ \citenamefont
  {Du}}]{x41}%
  \BibitemOpen
  \bibfield  {author} {\bibinfo {author} {\bibfnamefont {C.}~\bibnamefont
  {Zhang}}, \bibinfo {author} {\bibfnamefont {Y.}~\bibnamefont {Nie}}, \bibinfo
  {author} {\bibfnamefont {S.}~\bibnamefont {Sanvito}},\ and\ \bibinfo {author}
  {\bibfnamefont {A.}~\bibnamefont {Du}},\ }\href@noop {} {\bibfield  {journal}
  {\bibinfo  {journal} {Nano Lett.}\ }\textbf {\bibinfo {volume} {19}},\
  \bibinfo {pages} {1366} (\bibinfo {year} {2019})}\BibitemShut {NoStop}%
\bibitem [{\citenamefont {Smoleński}\ \emph {et~al.}(2016)\citenamefont
  {Smoleński}, \citenamefont {Goryca}, \citenamefont {Koperski}, \citenamefont
  {Faugeras}, \citenamefont {Kazimierczuk}, \citenamefont {Bogucki},
  \citenamefont {Nogajewski}, \citenamefont {Kossacki},\ and\ \citenamefont
  {Potemski}}]{x42}%
  \BibitemOpen
  \bibfield  {author} {\bibinfo {author} {\bibfnamefont {T.}~\bibnamefont
  {Smoleński}}, \bibinfo {author} {\bibfnamefont {M.}~\bibnamefont {Goryca}},
  \bibinfo {author} {\bibfnamefont {M.}~\bibnamefont {Koperski}}, \bibinfo
  {author} {\bibfnamefont {C.}~\bibnamefont {Faugeras}}, \bibinfo {author}
  {\bibfnamefont {T.}~\bibnamefont {Kazimierczuk}}, \bibinfo {author}
  {\bibfnamefont {A.}~\bibnamefont {Bogucki}}, \bibinfo {author} {\bibfnamefont
  {K.}~\bibnamefont {Nogajewski}}, \bibinfo {author} {\bibfnamefont
  {P.}~\bibnamefont {Kossacki}},\ and\ \bibinfo {author} {\bibfnamefont
  {M.}~\bibnamefont {Potemski}},\ }\href@noop {} {\bibfield  {journal}
  {\bibinfo  {journal} {Phys. Rev. X}\ }\textbf {\bibinfo {volume} {6}},\
  \bibinfo {pages} {021024} (\bibinfo {year} {2016})}\BibitemShut {NoStop}%
\bibitem [{\citenamefont {Xiao}\ \emph {et~al.}(2012)\citenamefont {Xiao},
  \citenamefont {Liu}, \citenamefont {Feng}, \citenamefont {Xu},\ and\
  \citenamefont {Yao}}]{x43}%
  \BibitemOpen
  \bibfield  {author} {\bibinfo {author} {\bibfnamefont {D.}~\bibnamefont
  {Xiao}}, \bibinfo {author} {\bibfnamefont {G.-B.}\ \bibnamefont {Liu}},
  \bibinfo {author} {\bibfnamefont {W.}~\bibnamefont {Feng}}, \bibinfo {author}
  {\bibfnamefont {X.}~\bibnamefont {Xu}},\ and\ \bibinfo {author}
  {\bibfnamefont {W.}~\bibnamefont {Yao}},\ }\href@noop {} {\bibfield
  {journal} {\bibinfo  {journal} {Phys. Rev. Lett.}\ }\textbf {\bibinfo
  {volume} {108}},\ \bibinfo {pages} {196802} (\bibinfo {year}
  {2012})}\BibitemShut {NoStop}%
\bibitem [{\citenamefont {Yao}\ \emph {et~al.}(2008)\citenamefont {Yao},
  \citenamefont {Xiao},\ and\ \citenamefont {Niu}}]{x44}%
  \BibitemOpen
  \bibfield  {author} {\bibinfo {author} {\bibfnamefont {W.}~\bibnamefont
  {Yao}}, \bibinfo {author} {\bibfnamefont {D.}~\bibnamefont {Xiao}},\ and\
  \bibinfo {author} {\bibfnamefont {Q.}~\bibnamefont {Niu}},\ }\href@noop {}
  {\bibfield  {journal} {\bibinfo  {journal} {Phys. Rev. B}\ }\textbf {\bibinfo
  {volume} {77}},\ \bibinfo {pages} {235406} (\bibinfo {year}
  {2008})}\BibitemShut {NoStop}%
\bibitem [{\citenamefont {Liu}\ \emph {et~al.}(2013)\citenamefont {Liu},
  \citenamefont {Shan}, \citenamefont {Yao}, \citenamefont {Yao},\ and\
  \citenamefont {Xiao}}]{x45}%
  \BibitemOpen
  \bibfield  {author} {\bibinfo {author} {\bibfnamefont {G.-B.}\ \bibnamefont
  {Liu}}, \bibinfo {author} {\bibfnamefont {W.-Y.}\ \bibnamefont {Shan}},
  \bibinfo {author} {\bibfnamefont {Y.}~\bibnamefont {Yao}}, \bibinfo {author}
  {\bibfnamefont {W.}~\bibnamefont {Yao}},\ and\ \bibinfo {author}
  {\bibfnamefont {D.}~\bibnamefont {Xiao}},\ }\href@noop {} {\bibfield
  {journal} {\bibinfo  {journal} {Phys. Rev. B}\ }\textbf {\bibinfo {volume}
  {88}},\ \bibinfo {pages} {085433} (\bibinfo {year} {2013})}\BibitemShut
  {NoStop}%
\bibitem [{\citenamefont {Ando}(1991)}]{x46}%
  \BibitemOpen
  \bibfield  {author} {\bibinfo {author} {\bibfnamefont {T.}~\bibnamefont
  {Ando}},\ }\href@noop {} {\bibfield  {journal} {\bibinfo  {journal} {Phys.
  Rev. B}\ }\textbf {\bibinfo {volume} {44}},\ \bibinfo {pages} {8017}
  (\bibinfo {year} {1991})}\BibitemShut {NoStop}%
\bibitem [{\citenamefont {Sekine}\ and\ \citenamefont
  {MacDonald}(2020)}]{Sekine102}%
  \BibitemOpen
  \bibfield  {author} {\bibinfo {author} {\bibfnamefont {A.}~\bibnamefont
  {Sekine}}\ and\ \bibinfo {author} {\bibfnamefont {A.~H.}\ \bibnamefont
  {MacDonald}},\ }\href@noop {} {\bibfield  {journal} {\bibinfo  {journal}
  {Phys. Rev. B}\ }\textbf {\bibinfo {volume} {102}},\ \bibinfo {pages}
  {155205} (\bibinfo {year} {2020})}\BibitemShut {NoStop}%
\bibitem [{\citenamefont {Klaiman}\ \emph {et~al.}(2007)\citenamefont
  {Klaiman}, \citenamefont {Moiseyev},\ and\ \citenamefont
  {Sadeghpour}}]{Klaiman}%
  \BibitemOpen
  \bibfield  {author} {\bibinfo {author} {\bibfnamefont {S.}~\bibnamefont
  {Klaiman}}, \bibinfo {author} {\bibfnamefont {N.}~\bibnamefont {Moiseyev}},\
  and\ \bibinfo {author} {\bibfnamefont {H.~R.}\ \bibnamefont {Sadeghpour}},\
  }\href@noop {} {\bibfield  {journal} {\bibinfo  {journal} {Phys. Rev. B}\
  }\textbf {\bibinfo {volume} {75}},\ \bibinfo {pages} {113305} (\bibinfo
  {year} {2007})}\BibitemShut {NoStop}%

\end{thebibliography}
\end{document}